\documentclass[nojss]{jss}
\usepackage{amsfonts}
\usepackage{amsmath}
\usepackage{bbm}
\usepackage{rotating}
\usepackage{booktabs}
\usepackage{multirow}
\def\qmo{``}
\def\qmc{''}
\def\qmcsp{'' }

\def\btheta{\mbox{\boldmath $\theta$}}
\def\beeta{\mbox{\boldmath $\eta$}}
\def\bF{\mathbf{F}}
\def\ba{\mathbf{a}}
\def\by{\mathbf{y}}
\def\bm{\mathbf{m}}
\def\bW{\mathbf{W}}
\def\bA{\mathbf{A}}
\def\bR{\mathbf{R}}
\def\bC{\mathbf{C}}
\newcommand{\indep}{\rotatebox[origin=c]{90}{$\models$}}

\author{Leopoldo Catania\\University of Rome, \qmo Tor Vergata\qmcsp \And
        Nima Nonejad\\ Aalborg University and CREATES}
\title{Dynamic Model Averaging for Practitioners in Economics and Finance: The \pkg{eDMA} Package}

\Plainauthor{Leopoldo Catania, Nima Nonejad} 
\Plaintitle{Dynamic Model Averaging for Practitioners in Economics and Finance: The eDMA Package} 
\Shorttitle{\pkg{eDMA}: Efficient Dynamic Model Averaging} 

\Abstract{
\cite{raftery_etal.2010} introduce an estimation technique, which they refer to as Dynamic
Model Averaging (DMA). In their application, DMA is used to predict the output strip thickness for a cold rolling mill,
where the output is measured with a time delay. Recently, DMA has also shown to be useful in macroeconomic and financial applications. In this paper, we present the \pkg{eDMA} package for DMA estimation implemented in \proglang{R}. The \pkg{eDMA} package is especially suited for practitioners in economics
and finance, where typically a large number of predictors are available. Our implementation is up to 133 times faster then a standard implementation using a single--core CPU. Thus, with the help of this package, practitioners are able to perform DMA on a standard PC without resorting to large clusters, which are not easily available to all researchers.
We demonstrate the usefulness of this package through simulation
experiments and an empirical application using quarterly U.S. inflation data.
}
\Plainkeywords{Dynamic model averaging, Multi core CPU, Parallel computing, R, OpenMP} 

\Address{
  Leopoldo Catania\\
  Department of Economics and Finance\\
  Faculty of Economics\\
  University of Rome, \qmo Tor Vergata\qmcsp\\
  Via Columbia, 2\\
  00133 Rome, Italy\\
  E-mail: \email{leopoldo.catania@uniroma2.it}\\

  Nima Nonejad\\
  Department of Mathematical Sciences\\
  Aalborg University and CREATES\\
  Fredrik Bajers Vej 7G\\
  9220, Aalborg East, Denmark\\
  E-mail: \email{nimanonejad@gmail.com}\\
}

\begin{document}

%
\section[Introduction]{Introduction}
\label{sec: Introduction}
Modeling and forecasting economic variables such as real GDP, inflation and equity premium
is of clear importance to researchers in economics and finance. For instance, forecasting inflation is
crucial for central banks with regards to conducting optimal monetary policy. Similarly, understanding and predicting equity premium is one of the most widely important topics discussed in financial economics as it has great implications on portfolio choice and risk management, see for instance \cite{dangl_halling.2012} among many others.

In order to obtain the best forecast as possible, practitioners often
try to take advantage of the many predictors available and
seek to combine the information from these predictors in an optimal
way, see \cite{stock_watson.1999}, \cite{stock_watson.2008} and \cite{groen_etal.2013} just to mention a few references. Recently, in the context of forecasting U.S. and UK inflation, \cite{koop_korobilis.2011} and \cite{koop_korobilis.2012}, implement a technique developed by \cite{raftery_etal.2010}, referred to as Dynamic
Model Averaging (DMA). The original purpose of the DMA introduced in \cite{raftery_etal.2010} is more oriented towards engineers. Particularly, their aim is to predict the output strip thickness for a cold rolling
mill, where the output is measured with a time delay. DMA consists of many time--varying coefficient regression models formed from all possible combinations of the predictors available to a practitioner. Moreover, besides allowing for time--variation in the regression coefficients, interpreting inclusion probabilities of each individual model, DMA also allows the relevant model set to change with time as well through a forgetting factor. This way, past model performance receives relatively less weight than current model performance and the estimation procedure adapts better to the incoming data. \cite{koop_korobilis.2011} and \cite{koop_korobilis.2012} argue that by slightly adjusting the original framework of \cite{raftery_etal.2010}, DMA can be useful in economic applications, especially inflation forecasting.\footnote{Specifically, \cite{koop_korobilis.2012} change the conditional volatility formula of \cite{raftery_etal.2010} arguing that the original formula is not suited for the analysis of economic data.} \cite{dangl_halling.2012} provide further suggestions on how to
improve DMA such that it can better adapt to the patterns typically observed in economic and financial data.
The aforementioned authors, also provide a useful variance decomposition scheme using the output from the estimation procedure. \cite{byrne_etal.2017}, among others, use the modifications proposed in \cite{dangl_halling.2012} to model currency exchange--rate behavior. We must also emphasize that DMA is not solely limited to these series and can be used in a wide range of economic applications such as: Forecasting realized volatility as well as house, oil and commodity prices.

However, from a practical point of view, designing an efficient DMA algorithm remains a challenging issue. As we demonstrate in Section~\ref{sec:ModifiedDMA}, DMA considers all possible combinations of predictors and forgetting factor values at each time--period. Typically, many candidate variables
are available and, as a consequence, it poses a limit given the computational facilities at hand, which for many practitioners typically consists of a standard 8 core CPU. In most cases, averaging over a relatively small number of model combinations (usually between 1000 to 3000) allows one to perform DMA using standard loops and software. However, handling larger number of combinations can quickly become very cumbersome and impose technical limits on the software at hand, especially with regards to memory consumption, see for example, \cite{koop_korobilis.2012}.
In order to deal with this issue,
\cite{onorante_raftery.2016} suggest a strategy that considers not the
whole model space, but rather a subset of models and dynamically optimizes
the choice of models at each point in time. However, \cite{onorante_raftery.2016} have to assume that models
do not change too fast over time, which is not an ideal assumption when dealing with financial and in some cases monthly economic data. Furthermore, it is not clear to us how one can incorporate the modifications suggested in \cite{dangl_halling.2012} within the framework of \cite{onorante_raftery.2016}.

In this paper, we introduce the \pkg{eDMA} package for \proglang{R} \citep{R.2015}, which efficiently implements a DMA procedure based on \cite{raftery_etal.2010} and \cite{dangl_halling.2012}. The routines in the \pkg{eDMA} package are principally written in \proglang{C++} using the \code{armadillo} library of \cite{sanderson.2010} and then made available in \proglang{R} exploiting the \pkg{Rcpp} and \pkg{RcppArmadillo} packages of \cite{Rcpp} and \cite{RcppArmadillo}, respectively. Furthermore, the \code{OpenMP} API \citep{openmp.2008} is used to speedup the computations when a shared memory multiple processors hardware is available, which, nowadays, is standard for the majority of commercial laptops. However, if the hardware does not have multiple processors, the \pkg{eDMA} package can still be used with the classical sequential CPU implementation. Our aim is to provide a package that can be used by a broad audience from different academic fields who are interested in implementing DMA in their research and obtain quantities such as: Inclusion probabilities, out--of--sample forecasts or to perform variance decomposition. Furthermore, our package enables practitioners, to perform DMA using a large number of predictors without needing to understand and possibly implement complex programming concepts such as \qmo how to efficiently allocate memory\qmc, or \qmo how to efficiently parallelize the computations\qmc.

It is also worth noting that, within the \proglang{R} environment, the \pkg{dma} package of \cite{dma} downloadable from CRAN can be used to perform the DMA of \cite{raftery_etal.2010}. However, \pkg{dma} has several weaknesses such as (i): It does not allow for the extensions mentioned in \cite{dangl_halling.2012}, which are important in the context of interpreting the amount of time--variation in the regression coefficients and performing a variance decomposition analysis, (ii): It is slow compared to the package introduced in this paper, (iii): It requires a very large amount of RAM when executed for moderately large applications, and (iv): It does not allow for parallel computing. We refer the reader interested in these aspects to Section \ref{sec:computational}, where we report a comparative analysis between \pkg{dma} and \pkg{eDMA} using simulated data. Moreover, \pkg{eDMA} permits us to also perform Bayesian Model Averaging (BMA) and Bayesian Model Selection (BMS) for linear regression models with constant coefficients implemented, for example, in the \proglang{R} packages \pkg{BMA} \citep{BMA} and \pkg{BMS} \citep{zeugner_feldkircher.2015}. At the same time, we obtain quantities such as: Posterior inclusion probabilities and average model size, which allow us to compare DMA (as well as Dynamic Model Selection, DMS) with BMA (BMS) with regards to model shrinkage and the magnitude of variation in the average model size.

The structure of this paper is as follows: Sections \ref{sec:Framework} and \ref{sec:ModifiedDMA} briefly introduce
DMA and its extensions. Section \ref{sec:ThePackage} presents
the technical aspects. Section \ref{sec:computational} provides an intuitive description of the challenges
that DMA posses from a computational point of view and proposes solutions. Section \ref{sec:EmpiricalApplication} provides an empirical application
to demonstrate the advantages of \pkg{eDMA} from a practical
point of view. Therefore, practitioners who are solely interested
on how to implement DMA using the \pkg{eDMA} package can skip Sections \ref{sec:Framework} and \ref{sec:ModifiedDMA}. Finally, Section \ref{sec:conclusion}
concludes.

\section[Framework]{Framework}
\label{sec:Framework}

\begin{figure}
\centering
\includegraphics[width=1\textwidth]{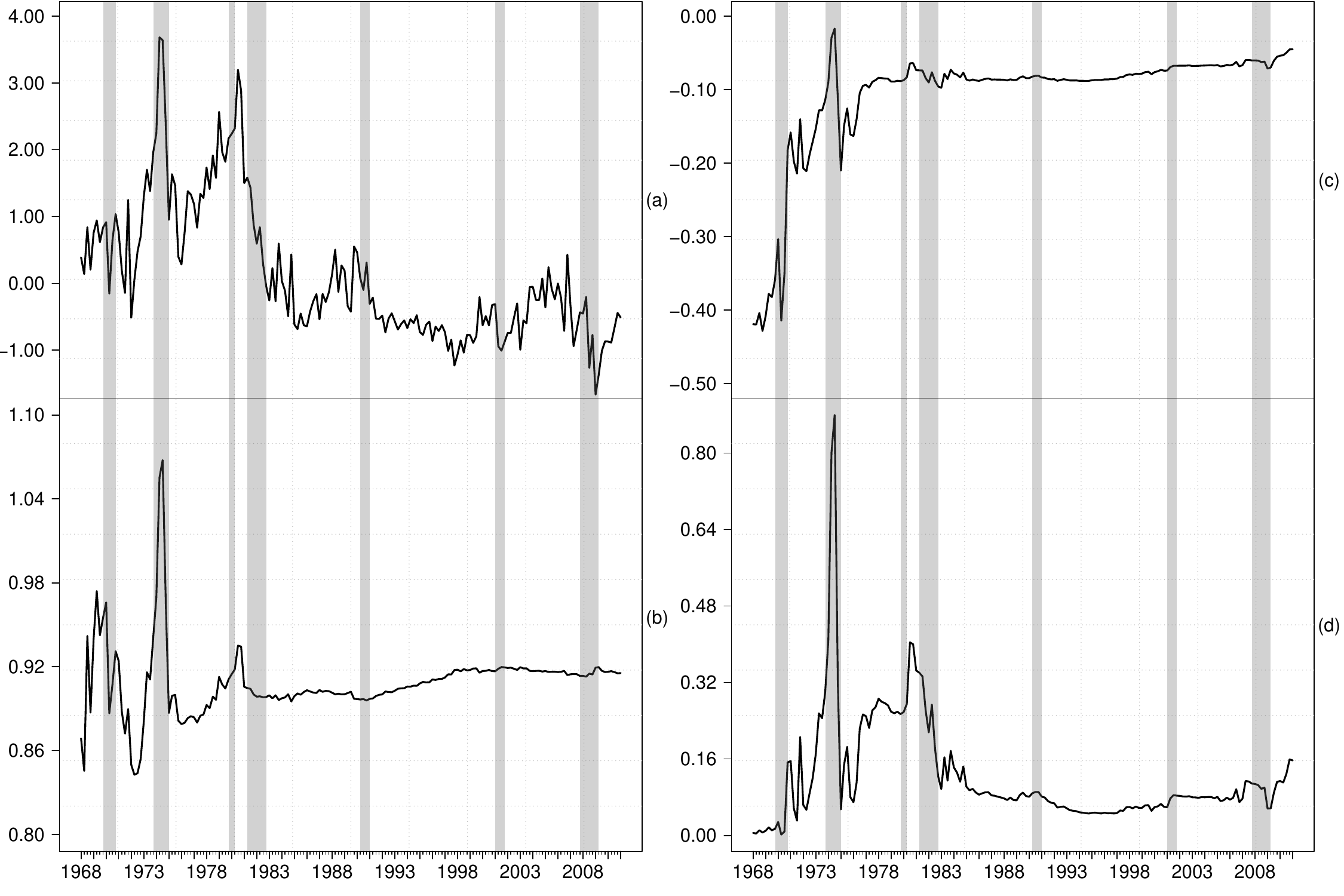}
\caption{Panel (a): Quarterly GDPDEF inflation from $1968$q1
to $2011$q2. Panel (b): Inflation persistence estimates from an
AR(2) model. Panel (c): Recursive OLS estimates of, $\theta_{4}$,
in $y_{t}=\theta_{1}+\theta_{2}y_{t-1}+\theta_{3}y_{t-2}+\theta_{4}\mathrm{UNEMP}_{t-1}+e_{t}$,
where $e_{t}\sim N\left(0,\sigma^{2}\right)$. Panel (d): Recursive $p$--values
for the null hypothesis of $\theta_{4}=0$ at the $5\%$ level. The
gray vertical bars indicate business cycle peaks, \emph{i.e.}, the point
at which an economic expansion transitions to a recession, based on
National Bureau of Economic Research (NBER) business cycle dating.}
\label{fig:Figure 0}
\end{figure}

Many forecasting applications are based on a model where the variable
of interest at time $t$, $y_{t}$, depends on exogenous predictors
and possibly lagged values of $y_{t}$ itself. For instance, in panel
(a) of Figure \ref{fig:Figure 0}, we plot the quarterly U.S. inflation
rate, $100\triangle\ln P_{t}$, where $P_{t}$ denotes the U.S. Gross
Domestic Product implicit price deflator (GDPDEF) from $1968$q1 to
$2011$q2. We then
recursively (\emph{i.e.}, using data up to time $t$) estimate an autoregressive model
of order 2, AR(2), of $y_{t}$ and report the sum of the autoregressive
coefficients, which can be considered as a basic measure of inflation persistence in Panel
(b). Our general conclusions from panels (a)--(b) are: Inflation is
persistent and generally tends to be higher during recessions than
tranquil periods. It does not appear to follow an identical cyclical
pattern either. For example, inflation increases less aggressively
towards the Great Recession of $2008$ than the corresponding downturns in the $1970$s,
$1980$s or the early $2000$s. Furthermore, even in this simple model,
we still observe some time--variation in the AR coefficients. We then extend the plain AR(2) model to also include the
lagged unemployment rate (UNEMP) as a regressor. This way, we end
up with a basic Philips curve. In panel (c), we report the recursive
OLS estimates of UNEMP and in panel (d), we report the recursive p--values associated to the null hypothesis that this estimate is equal to zero. Panel (d) shows that the unemployment rate in some periods seems to be a useful predictor of inflation.

Results from panels (a)--(d) of Figure \ref{fig:Figure 0} suggest
that two channels can potentially help to improve the accuracy of
inflation forecasts, (i): Incorporating time--variation in the regression
coefficients. (ii): Augmenting the AR model with exogenous predictors
that can capture information beyond that already contained in lagged
values of $y_{t}$. Thus, in many economic applications, we eventually
end up with a model such as:

\begin{equation}
y_{t} = \theta_{1t}+\theta_{2t}y_{t-1}+\theta_{3t}y_{t-2}+\theta_{4t}x_{t-1}+...+\theta_{nt}z_{t-1}+\varepsilon_{t},\quad\varepsilon_{t}\sim N\left(0,V_{t}\right).\label{eq:1.1}
\end{equation}

Obviously, $n$ can be large and as a consequence, we may have to deal with a very large number of model combinations. For example, if our set of models is defined by whether
each of the $n$ potential predictors is included or excluded, then
we can have as high as $k=2^{n}$ model combinations to consider, which
raises substantive challenges for model selection. This aspect is
referred to as \qmo model uncertainty\qmc, \emph{i.e.}, the uncertainty that
a practitioner faces in choosing the correct combination of predictors.
It is important to note, that discarding this aspect can have severe consequences
on out--of--sample results. This is due to the fact that, simply adding additional predictors in
our model without designing an optimal model selection strategy, can deteriorate out--of--sample performance due to the bias--variance trade--off (the
additional reduction in bias afforded by including additional predictors
does not offset the increased forecast variance related to the more
heavily parameterized model). Besides model uncertainty, a practitioner
also faces uncertainty regarding the nature of time--variation in the
regression coefficients, \emph{i.e.}, \qmo parameter uncertainty\qmc. Underestimating or overestimating the magnitude of time--variation in the regression coefficients also has important consequences as our model adapts either too slowly or too quickly to new data, generating either too rigid or too volatile forecasts.
The DMA methodology provides an optimal way to deal with these sources of
uncertainty. Moreover, it is simple, parsimonious and allows us to evaluate
out--of--sample forecasts based on a large set of model combinations
in real--time (no need to condition on the full sample at time $t$)
without resorting to simulation.

To provide more details on the underlying mechanism of DMA, we start by assuming that any combination of the elements on the right--hand--side
of (\ref{eq:1.1}) can be expressed as a Dynamic Linear Models (DLM),
see \cite{west_harrison.1999} and \cite{raftery_etal.2010}. Particularly,
let $\bF_{t}^{\left(i\right)}$denote a $p\times 1$ vector based on a given combination
of our total predictors, $\bF_{t}=\left(1, y_{t-1},y_{t-2},x_{t-1},...,z_{t-1}\right)^\top$.
Then, we can express our $i$--th DLM as:

\begin{align}
y_{t} & =  \bF_{t}^{\left(i\right)\prime}\btheta_{t}^{\left(i\right)}+\varepsilon_{t}^{\left(i\right)},\quad\varepsilon_{t}^{\left(i\right)}\sim N\left(0,V_{t}^{\left(i\right)}\right)\label{eq:2.1}\\
\btheta_{t}^{\left(i\right)} & =  \btheta_{t-1}^{\left(i\right)}+\beeta_{t}^{\left(i\right)},\quad\beeta_{t}^{\left(i\right)}\sim N\left(0,\bW_{t}^{\left(i\right)}\right),\label{eq:2.2}
\end{align}

where the $p\times 1$ vector of time--varying regression coefficients, $\btheta_{t}^{\left(i\right)} = \left(\theta_{1t}^{(i)}, \dots, \theta_{pt}^{(i)}\right)^\top$, evolves according to (\ref{eq:2.2}) and determines the impact of $\bF^{(i)}_t$ on $y_t$.
Note, we do not assume any
systematic movements in $\btheta_{t}^{\left(i\right)}$. On the contrary,
we consider changes in $\btheta_{t}^{\left(i\right)}$ as unpredictable.\footnote{See \cite{dangl_halling.2012} and \cite{koop_korobilis.2012} for a similar
model specification.}

The conditional variances, $V^{\left(i\right)}_{t}$ and $\bW^{\left(i\right)}_{t}$, are unknown quantities associated with the observational equation, \eqref{eq:2.1}, and the state equation, \eqref{eq:2.2}. Obviously, when $\bW^{\left(i\right)}_{t}=\boldsymbol{0}$ for $t=1,\dots,T$, then $\btheta^{\left(i\right)}_{t}$ is constant over time. Thus, \eqref{eq:2.1}--\eqref{eq:2.2} nests the specification of constant regression coefficients. For $\bW^{\left(i\right)}_{t}\neq\boldsymbol{0}$, $\btheta^{\left(i\right)}_{t}$ varies according to Equation~\ref{eq:2.2}. However, this does not mean that $\btheta^{(i)}_t$ needs to change at every time period. For instance, we can easily have periods where $\bW^{(i)}_t=\boldsymbol{0}$ and thus $\btheta^{(i)}_t=\btheta^{(i)}_{t-1}$. Ultimately, the nature of time variation in the regression coefficients is dependent on the data at hand.\footnote{We model time--variation in $\btheta_t^{(i)}$ through a forgetting factor, $\delta$, see below for more details. Moreover, we show that the recursive updating of the forgetting factor based on the predictive likelihood, avoids any unreasonable behavior of $\btheta_t^{(i)}$ even though, we do not specifically put any structure on $\btheta^{(i)}_t$. We also refer the reader to Appendix A.3 of \cite{dangl_halling.2012}, where it is shown that \eqref{eq:2.2} outperforms the autoregressive structured counterpart.}

In DMA, we consider a total of $k=2^{n}-1$ possible combinations of the predictors at each point in time while contemporaneously assuming that $\btheta^{(i)}_t$ can evolve over time.\footnote{The model $y_t=\varepsilon_t$ is \textbf{not} considered in the universe of models, see also \cite{dangl_halling.2012}.} DMA then averages forecasts across the different combinations using a recursive updating scheme based on the predictive likelihood. The predictive likelihood measures the ability of a model to predict $y_t$, thus making it
the central quantity of interest for model evaluation. Models containing important combinations of predictors receive high predictive likelihood values, which means that they obtain relatively higher posterior weights in the averaging process. Besides averaging, we can also use the forecasts of the model receiving the highest probability among all model combinations considered at each point in time. In this case, we are performing Dynamic Model Selection (DMS), see also \cite{koop_korobilis.2012}.

As indicated in \eqref{eq:2.2}, we must specify $\bW^{(i)}_t, i=1,\dots,k$. Obviously, this task can be very daunting if we were to specify $\bW^{(i)}_t$ for each of the total $k$ models. However, DMA avoids the difficult task of specifying $\bW^{\left(i\right)}_{t}$ for each individual model relying on a forgetting factor, $0<\delta\leq1$. This in turn simplifies things greatly from a practical point of view as instead of working with many parameters, we only need to worry about $\delta$. Now, we briefly explain how this mechanism works. We start by defining the variables of the Kalman recursions for the $i$--th model as follows: (i): $\bR^{\left(i\right)}_{t}$, the prediction variance of $\btheta^{\left(i\right)}_t$ (see Equation \ref{eq:A.1} in Appendix \ref{sec:app1} at the end of paper), (ii): $\bC^{\left(i\right)}_{t}$, the estimator for the covariance matrix of $\btheta^{\left(i\right)}_t$, (see Equation \ref{eq:A.3}), and (iii): $S^{\left(i\right)}_{t}$, the estimator of the observational variance.
Then, using $\delta$, we can rewrite $\bR_t^{(i)}=\bC^{(i)}_{t-1}+\bW^{(i)}_t$ in Appendix \ref{sec:app1} as $\bR^{\left(i\right)}_{t}=\delta^{-1}\bC^{(i)}_{t-1}$, indicating that there is a relationship between $\bW^{\left(i\right)}_{t}$ and $\delta$, which is given as $\bW^{\left(i\right)}_{t}=\left(1-\delta\right)/\delta \bC^{\left(i\right)}_{t-1}$. In other words, the loss of information is proportional to the covariance of the state parameters, $\bC^{(i)}_{t}$.
Clearly, we can control the magnitude of the shocks that impact $\btheta^{\left(i\right)}_{t}$
by adjusting $\delta$ instead of directly estimating $\bW^{(i)}_t$. Accordingly, $\delta=1$ corresponds to
$\bW^{\left(i\right)}_{t}=\boldsymbol{0}$, which means that $\btheta_{t}^{(i)}$ equals its value at time $t-1$. For $\delta<1$, we introduce time--variation in $\btheta^{\left(i\right)}_{t}$. For instance, when $\delta=0.99$, in the context quarterly data, observations five years ago receive approximately $80\%$ as much weight as last period's observation, which corresponds to gradual time--variation in $\btheta^{(i)}_t$. When $\delta=0.95$, observations $20$ periods ago receive only about $35\%$ as much weight as last period's observations, suggesting that a relatively larger shock hits the regression coefficients. Evidently, while this renders the model more
flexible to adapt to changes in the data, the increased variability
in $\btheta^{\left(i\right)}_{t}$ also results in higher prediction variance. Thus, estimating \eqref{eq:2.1}--\eqref{eq:2.2} depends not only on the choice of the predictors in $\bF^{(i)}_t$ but also the choice of $\delta$.

Conditional on $\delta$, the DMA probability of model $M_{i}$
conditional on the current information set at time $t$, $\mathcal{F}_{t}$,
is then defined as:

\begin{equation}
p\left(M_{i}\mid\mathcal{F}_{t}\right) = \frac{p\left(y_{t}\mid M_{i},\mathcal{F}_{t-1}\right)p\left(M_{i}\mid\mathcal{F}_{t-1}\right)}{\sum_{l=1}^{k}p\left(y_{t}\mid M_{l},\mathcal{F}_{t-1}\right)p\left(M_{l}\mid\mathcal{F}_{t-1}\right)},\nonumber
\end{equation}

where $p\left(y_{t}\mid M_{i},\mathcal{F}_{t-1}\right)$ is the predictive
likelihood of model $M_{i}$ evaluated at $y_{t}$,  $p\left(M_{i}\mid\mathcal{F}_{t-1}\right)=p\left(M_{i}\mid\mathcal{F}_{t-1}\right)^{\alpha}/\Sigma_{l=1}^{k}p\left(M_{l}\mid\mathcal{F}_{t-1}\right)^{\alpha}$
where $\mbox{0<\ensuremath{\alpha}}\leq1$ is the forgetting factor
for the entire model chosen by the practitioner and $p(M_i\vert\mathcal{F}_{t-1})$ is the model probability at time $t-1$. The forgetting factor parameter, $\alpha$, induces time--variation in the entire model set. Clearly, the lower the value
of $\alpha$, the lesser weight is given to past performance. \cite{raftery_etal.2010} and \cite{koop_korobilis.2012} recommend setting $\alpha$
close to one. \cite{dangl_halling.2012}, on the other hand, fix $\alpha$ at $1$.

Finally, we must also determine a way to model the evolution of $V_{t}^{\left(i\right)}, i=1,\dots,k$. Here, we have two options, which we go into more details below, see point (c).
Thus in order to initialize the DMA recursions, a practitioner must:

\begin{itemize}
\item[(a):] Consider the number of predictors. Typically, in economic applications, we use exogenous variables as well as lagged values of $y_t$ as predictors. For instance,
in the context of forecasting quarterly inflation, besides considering predictors such as unemployment and T--bill rates, \cite{koop_korobilis.2012} also consider the first three lags of $y_{t}$ as predictors.

\item[(b):] Choose $\delta$ and $\alpha$. In many applications
$\alpha\in\left\{ 0.98,0.99,1\right\}$ works well and generally results do not change drastically across different values of $\alpha$\footnote{Recently, in the context of binary regressions, \cite{mccormick_etal.2012} suggest a technique where one can model $\alpha$ as time--varying.}. On the other
hand, as previously mentioned, we often find that the choice of $\delta$ is more important.
\cite{koop_korobilis.2012} fix $\delta$ at $\{0.95,0.98,0.99,1.00\}$ and run DMA using each of these values. They find that results differ considerably in terms
of out--of--sample forecasts. Evidently, in many economic applications,
it is plausible that $\delta$ would indeed be time--varying.
For instance, it is plausible to expect that $\delta$ is relatively low in
recessions or periods of market turmoil (where there is considerable
time--variation in $\btheta^{\left(i\right)}_{t}$). Conversely, $\delta$ is ought
to be close to $1.00$ during tranquil periods, where basically nothing changes. \cite{dangl_halling.2012} propose an elegant solution to this problem by considering a grid of values for $\delta$ and incorporate this in the DMA setting by averaging over all possible combinations of the predictors as well as the corresponding grid of $\delta$. At the same time, this strategy means that we avoid any unreasonable behavior of $\btheta^{(i)}_t$ as $\delta$ values incompatible with the data (and of course results in bad behavior on $\btheta^{(i)}_t$) do not receive a weight in the averaging process. Furthermore, this procedure can also be used to obtain more information from the data through a variance decomposition scheme, see below for more details.
\item[(c):] Evolution of $V_{t}^{\left(i\right)}$: We can make things easy
for conjugate analysis by assuming that $V_{t}^{\left(i\right)}=V^{\left(i\right)}$
for all $t$. At time $t=0$, we specify a Normal prior on $\btheta_{0}^{\left(i\right)}$
and a Inverted--gamma prior on $V^{\left(i\right)}$, \emph{i.e.}, $V^{\left(i\right)}\vert\mathcal{F}_{0}\sim \mathcal{IG}\left(\frac{1}{2},\frac{1}{2}S_{0}^{\left(i\right)}\right)$,
where $\mathcal{IG}\left(\frac{v}{2},\frac{\kappa}{2}\right)$ stands for the
Inverted--gamma distribution with scale, $v$, and shape $\kappa$,
see also \cite{prado_west.2010}. Then, the posterior of $V^{(i)}$ follows an $\mathcal{IG}$ distribution with parameters, $S^{(i)}_t$, $n^{(i)}_t$, where the time $t$ point estimate of $V^{(i)}$, $S^{(i)}_t$, is given as

\begin{equation}
S_{t}^{\left(i\right)} = S_{t-1}^{\left(i\right)}+\frac{S_{t-1}^{\left(i\right)}}{n_{t}^{\left(i\right)}}\left(\frac{e_{t}^{2\left(i\right)}}{Q_{t}^{\left(i\right)}}-1\right),\nonumber
\end{equation}

$n^{(i)}_{t} = n^{(i)}_{t-1} + 1$, $e^{(i)}_{t}$ and $Q^{(i)}_{t}$ are given in Appendix \ref{sec:app1} and \cite{prado_west.2010}. Clearly, $S^{(i)}_t$ approaches to a constant level as $n^{(i)}_t$ increases. More importantly, under these assumptions,
we find that, when we integrate the conditional density of $y_{t}$
over the values of $\btheta_{t}^{\left(i\right)}$ and $V^{\left(i\right)}$,
the corresponding predictive density has a closed--form solution given
by, $\mathcal{T}_{n_{t}^{\left(i\right)}}\left(\hat{y}_{t}^{\left(i\right)},Q_{t}^{\left(i\right)}\right)$,
where $\mathcal{T}_{n_{t}^{\left(i\right)}}$ stands for the Student's t--distribution
with $n_{t}^{\left(i\right)}$ degrees--of--freedom, mean and scale
given by $\hat{y}_{t}^{\left(i\right)} = \bF^{(i)^\top}_t \ba^{(i)}_{t-1}$ and $Q_{t}^{\left(i\right)}$, see Appendix \ref{sec:app1} for more details.
\end{itemize}

However, in many applications, allowing for time--variation in the conditional error variance
better suits our underlying economic assumptions. Therefore, we follow \cite{prado_west.2010} and in a similar fashion as for $\bW_{t}^{\left(i\right)}$
adopt a discount factor to induce time--variation in $V_{t}^{\left(i\right)}$.
Particularly, we do this by imposing a forgetting factor, $0<\beta\leq1$, which enters the
scale and the shape parameters of the Inverted--gamma distribution, such that $n_{t}^{\left(i\right)} = \beta n_{t-1}^{\left(i\right)} + 1$. This way, $V_{t}^{\left(i\right)}$
is updated according to new data, forgetting past information to reflect
changes in volatility. This approach means that, if $\beta < 1$, the time $t$ estimate
of $V_{t}^{\left(i\right)}$ is given as:

\begin{equation}\label{eq:beta}
S_{t}^{\left(i\right)} = \left(1-\beta\right)\sum_{s=0}^{t-1}\beta^{s}\left(\frac{e_{t-s}^{2\left(i\right)}S_{t-s-1}^{\left(i\right)}}{Q_{t-s}^{\left(i\right)}}\right).
\end{equation}

In other words, $V_{t}^{\left(i\right)}$ has form of an exponentially
weighted moving average (EWMA) and older data are further discounted
as time progresses. When $\beta=1$, then we recover $V_{t}^{\left(i\right)}=V^{\left(i\right)}$\footnote{We would like to thank an anonymous referee for this suggestion.}.
This extension obviously requires the practitioner to also consider a value
for $\beta$. By experimenting with small models based primarily on simulated data, we
find that $\delta$ and $\beta$ in many ways are intertwined, in
the sense that we can recover the same magnitudes of variation in
$\btheta_{t}^{\left(i\right)}$ using different values of $\delta$
and $\beta$. For example, when we fix $\beta$ close to $1$ (below $1$, say $0.95$),
we find that a relatively lower (higher) value of $\delta$ is needed
to recover the fundamental dynamics in the regression coefficients. This is understandable
as allowing for variation in the conditional variance takes always
some dynamics from the regression coefficients, whereas more dynamics
in $\btheta_{t}^{\left(i\right)}$ are required in order to compensate
for the possible lack of time--variation in $V_{t}^{\left(i\right)}$. Overall,
our conclusion is that if a practitioner chooses to fix $\beta<1$, then it is best to fix $\delta$ close to $1$,
say at $0.96$, which is also to the value used by \cite{riskmetrics.1996}.
This way, we maintain a parsimonious model structure and allow for time--variation in $V^{(i)}_t$. More importantly, we are
never in doubt whether we are under (over) estimating the true magnitude
of variation in $\btheta_{t}^{\left(i\right)}$\footnote{We observe the same phenomena when we allow $\alpha$ to vary with $\delta$. Overall, our conclusion is that it is best to use (c) and fix $\alpha$ close to $0.99$ for monthly and quarterly data. However, if a practitioner wishes to set $\beta<1$, then we generally recommend $\beta>0.96$ and $\alpha=0.99$.}.

\section[Modified DMA]{Modified DMA}
\label{sec:ModifiedDMA}
Below, we present the DMA algorithm modified to incorporate the extensions mentioned in Section \ref{sec:Framework}. Let $M_{i}$ denote a model containing a specific set of predictors
chosen from a set of $k=2^n-1$ candidates and $\delta_{j}$
denotes a specific forgetting factor value chosen from a pre--specified grid of values, $\{\delta_1,\dots,\delta_d\}$. The total posterior density of model $M_{i}$ and forgetting factor value
$\delta_{j}$ at time $t$, $p\left(M_{i},\delta_{j}\vert\mathcal{F}_{t}\right)$,
is then given as
\begin{align}
p\left(M_{i},\delta_{j}\vert\mathcal{F}_{t}\right) & =  p\left(M_{i}\vert\delta_{j},\mathcal{F}_{t}\right)p\left(\delta_{j}\vert\mathcal{F}_{t}\right).\nonumber
\end{align}
In order to obtain $p\left(M_{i}\vert\mathcal{F}_{t}\right)$ we can use the relation
\begin{align}
p\left(M_{i}\vert\mathcal{F}_{t}\right) & =  \sum_{j=1}^{d}p\left(M_{i}\vert\delta_{j},\mathcal{F}_{t}\right)p\left(\delta_{j}\vert\mathcal{F}_{t}\right).\label{eq:3.2}
\end{align}
The term, $p\left(M_{i}\vert\delta_{j},\mathcal{F}_{t}\right)$, in Equation~\ref{eq:3.2} is given as
\begin{align}
p\left(M_{i}\vert\delta_{j},\mathcal{F}_{t}\right) & =  \frac{p\left(y_{t}\vert M_{i},\delta_{j},\mathcal{F}_{t-1}\right)p\left(M_{i}\vert\delta_{j},\mathcal{F}_{t-1}\right)}{\sum_{l=1}^{k}p\left(y_{t}\vert M_{l},\delta_{j},\mathcal{F}_{t-1}\right)p\left(M_{l}\vert\delta_{j},\mathcal{F}_{t-1}\right)}\label{eq:3.3}
\end{align}
where
\begin{align}
p\left(M_{i}\vert\delta_{j},\mathcal{F}_{t-1}\right) & =  \frac{p\left(M_{i}\vert\delta_{j},\mathcal{F}_{t-1}\right)^{\alpha}}{\sum_{l=1}^{k}p\left(M_{l}\vert\delta_{j},\mathcal{F}_{t-1}\right)^{\alpha}}.\label{eq:3.4}
\end{align}
The second term on the right--hand side of Equation~\ref{eq:3.2} is given as
\begin{align}\label{eq:delta}
p\left(\delta_{j}\vert\mathcal{F}_{t}\right) & =  \frac{p\left(y_{t}\vert\delta_{j},\mathcal{F}_{t-1}\right)p\left(\delta_{j}\vert\mathcal{F}_{t-1}\right)}{\sum_{l=1}^{d}p\left(y_{t}\vert\delta_{l},\mathcal{F}_{t-1}\right)p\left(\delta_{l}\vert\mathcal{F}_{t-1}\right)}.
\end{align}
where
\begin{align}
p\left(\delta_{j}\vert\mathcal{F}_{t-1}\right) & =  \frac{p\left(\delta_{j}\vert\mathcal{F}_{t-1}\right)^{\alpha}}{\sum_{l=1}^{d}p\left(\delta_{l}\vert\mathcal{F}_{t-1}\right)^{\alpha}}.\nonumber
\end{align}
Typically, $p\left(M_{i},\delta_{j}\vert\mathcal{F}_{0}\right)=1/(d\cdot k)$
such that, initially, all model combinations and degrees of time--variation
are equally likely. Thereafter, as a new observation arrives, model probabilities are updated using
the above recursions.
\subsection[Using the output from DMA]{Using the output from DMA}
For practitioners, the most interesting output from DMA are:
\begin{itemize}
\item[(i)] The predictive mean of $y_{t+1}$ conditional on $\mathcal{F}_{t}$, denoted by $\hat y_{t+1}$.
This is simply an average of each of the individual model predictive
means. That is
\begin{align}
\hat y_{t+1} & =  \sum_{j=1}^{d}\E\left[y_{t+1}^{\left(j\right)}\vert\mathcal{F}_{t}\right]p\left(\delta_{j}\vert\mathcal{F}_{t}\right),\label{eq:4.5}
\end{align}
where
\begin{align}
\E\left[y_{t+1}^{\left(j\right)}\vert\mathcal{F}_{t}\right] & =  \sum_{i=1}^{k}\E\left[y_{i,t+1}^{\left(j\right)}\vert\mathcal{F}_{t}\right]p\left(M_{i}\vert\delta_{j},\mathcal{F}_{t}\right).\nonumber
\end{align}
The formulas for the predictive density are given as
\begin{align}\label{eq:preddens}
p\left(y_{t+1}\vert\mathcal{F}_{t}\right) = & \sum_{j=1}^{d}p(y_{t+1}^{(j)}\vert\mathcal{F}_t)p(\delta_{j}\vert\mathcal{F}_{t}),
\end{align}
where
\begin{align}
p(y_{t+1}^{(j)}\vert\mathcal{F}_{t}) = & \sum_{i=1}^{k}p(y_{i,t+1}^{(j)}\vert\mathcal{F}_{t})p(M_{i}\vert\delta_{j},\mathcal{F}_{t}).\nonumber
\end{align}
Besides averaging over the individual predictive means/densities, we can simply choose the predictive mean/density associated with the model with the highest posterior probability. Henceforth, we label this as Dynamic Model Selection (DMS), see also \cite{koop_korobilis.2012}. When, $\delta$, $\beta$ and $\alpha$ are all fixed at 1, we have Bayesian model averaging \cite[BMA, see][]{raftery.1995} and Bayesian Model Selection (BMS) based on exact predictive likelihood, see for instance \cite{zeugner_feldkircher.2015}.\footnote{\cite{zeugner_feldkircher.2015} also implement BMA using the MC$^3$ algorithm relying on Markov Chain Monte Carlo (MCMC) techniques. However, their framework does not allow for time--variation in the regression coefficients nor model size.}
\item[(ii)] Quantities such as the expected size, $\E\left[Size_{t}\right]=\Sigma_{i=1}^{k}Size^{(i)}p\left(M_{i}\vert\mathcal{F}_{t}\right)$,
where $Size^{(i)}$ be the number of predictors in model
$i$. This quantity reveals the average number of predictors in the DMA, see \cite{koop_korobilis.2012}. Similarly, we can compute the number of predictors for the model with the highest posterior probability, \eqref{eq:3.2}, at each point in time, which give the optimal model size at time $t$.
\item[(iii)] Posterior inclusion probabilities for the predictors. That is,
at each $t$, we calculate $\sum_{i=1}^{k}1_{\left(i\subset m\right)}p\left(M_{i}\vert\mathcal{F}_{t}\right)$,
where $1_{\left(i\subset m\right)}$ is an indicator function taking
the value of either $0$ or $1$ and $m$, $m=1,\dots,n$, is the $m$th predictor. We can also report the highest posterior model probability or the sum of the top $10\%$ model probabilities among all model combinations after the effect of $\delta$ is integrated out. This information can be used to determine if there is a group or an individual model that obtains relatively high posterior probability.
\item[(iv)] Posterior weighted average of $\delta$ at each point in time that is $\sum_{j=1}^{d}\delta_j p\left(\delta_j\vert\mathcal{F}_t\right)$,  for $t=1,\dots,T$.
\item[(v)] Posterior weighted average estimates of $\btheta_{t}$ for DMA
\begin{align}\label{eq:theta_post}
\E[\btheta_{t}\vert\mathcal{F}_{t}] & =  \sum_{j=1}^{d}\E[\btheta_{t}^{(j)}\vert\mathcal{F}_{t}]p(\delta_{j}\vert\mathcal{F}_{t}),
\end{align}
where
\begin{align}
\E[\btheta_{t}^{(j)}\vert\mathcal{F}_{t}] & =  \sum_{i=1}^{k}\E[\btheta_{i,t}^{(j)}\vert\mathcal{F}_{t}]p(M_{i}\vert\delta_{j},\mathcal{F}_{t}).\nonumber
\end{align}

\item[(vi)] Variance decomposition of the data, $\textrm{Var}\left(y_{t+1}\vert\mathcal{F}_t\right)$,
 decomposed into:

\begin{align}\label{eq:varDec}
\VAR\left(y_{t+1}\vert\mathcal{F}_t\right) &= \mathrm{Obs}_{t+1} + \mathrm{Coeff}_{t+1} + \mathrm{Mod}_{t+1} + \mathrm{TVP}_{t+1}
\end{align}

where:

\begin{align}
\mathrm{Obs}_{t+1} & =  \sum_{j=1}^{d}\left[\sum_{i=1}^{k}\left(S_{t}\vert M_{i},\delta_{j},\mathcal{F}_{t}\right)p\left(M_{i}\vert\delta_{j},\mathcal{F}_{t}\right)\right]p\left(\delta_{j}\vert\mathcal{F}_{t}\right),\nonumber\\
\mathrm{Coeff}_{t+1} & = \sum_{j=1}^{d}\left[\sum_{i=1}^{k}\left(\bF_{t}^\top\bR_{t}\bF_{t}\vert M_{i},\delta_{j},\mathcal{F}_{t}\right)p\left(M_{i}\vert\delta_{j},\mathcal{F}_{t}\right)\right]p\left(\delta_{j}\vert\mathcal{F}_{t}\right),\nonumber\\
\mathrm{Mod}_{t+1} & = \sum_{j=1}^{d}\left[\sum_{i=1}^{k}\left(\hat{y}_{i,t+1}^{\left(j\right)}-\hat{y}_{t+1}^{\left(j\right)}\right)^{2}p\left(M_{i}\vert\delta_{j},\mathcal{F}_{t}\right)\right]p\left(\delta_{j}\vert\mathcal{F}_{t}\right),\nonumber\\
\mathrm{TVP}_{t+1} & = \sum_{j=1}^{d}\left(\hat{y}_{t+1}^{\left(j\right)}-\hat{y}_{t+1}\right)^{2}p\left(\delta_{j}\vert\mathcal{F}_{t}\right).
\end{align}

The first term is the observational variance, Obs. The remaining terms
are: Variance due to errors in the estimation of the coefficients,
Coeff, variance due to uncertainty with respect to the choice
of the predictors, Mod, and variance due to uncertainty with
respect to the choice of the degree of time--variation in the regression
coefficients, TVP, see \cite{dangl_halling.2012} for more details.
\end{itemize}

\section[The eDMA package for R]{The \pkg{eDMA} package for \proglang{R}}

\label{sec:ThePackage}
The \pkg{eDMA} package for \proglang{R} offers an integrated environment for practitioners in economics and finance to perform our DMA algorithm. It is principally written in \proglang{C++}, exploiting the \pkg{armadillo} library of \cite{sanderson.2010} to speed up computations. The relevant functions are then made available in \proglang{R} through the \pkg{Rcpp} and \pkg{RcppArmadillo} packages of \cite{Rcpp} and \cite{RcppArmadillo}, respectively. It also makes use of the \code{OpenMP} API \citep{openmp.2008} to parallelize part of the routines needed to perform DMA. Furthermore, multiple processors are automatically used if supported by the hardware, however, as will be discussed later, the user is also free to manage the level of resources used by the program.

The \pkg{eDMA} package is written using the S4 object oriented language, meaning that classes and methods are available in the code. Specifically, \proglang{R} users will find common methods such as \code{plot()}, \code{show()}, \code{as.data.frame()}, \code{coef()} and \code{residuals()}, among others, in order to visualise the output of DMA and extract estimated quantities.

The \pkg{eDMA} package is available from CRAN at \url{https://cran.r-project.org/web/packages/eDMA/index.html} and can be installed using the command:

\begin{CodeChunk}
\begin{CodeInput}
R> install.packages("eDMA")
\end{CodeInput}
\end{CodeChunk}

Once the package is correctly installed and loaded, the user faces one function named \code{DMA()} to perform DMA. The \code{DMA()} function then accepts a series of arguments and returns an object of the class \code{DMA} which comes with several methods, see Section \ref{sec:methods}. The arguments the \code{DMA()} function accepts are:
\begin{itemize}
  \item \code{formula}: An object of class \code{formula} (or one that can be coerced to that class): A symbolic description of the model to be fitted. The formula should include all the predictors one chooses to use. The inclusion of the constant term follows the usual \proglang{R} practice, \emph{i.e.}, it is included by default and can be removed if necessary. For instance, in order to model \code{y ~ x}, however, without the constant, we can write for example, \code{y ~ x - 1}, see \code{help(formula)}. This implementation follows the common practice for \proglang{R} users, see \emph{e.g.}, the \pkg{plm} package of \cite{plm}.
  \item \code{data}: A \code{data.frame} (or object coercible by \code{as.data.frame()} to a \code{data.frame}) containing the variables in the model. If \code{data} is an object of the class \code{ts}, \code{zoo} or \code{xts}, then the time information is used in the graphical representation of the results as well as for the estimated quantities. The dimension of \code{data} is $T\times \left(1 + n\right)$, containing at each row, the dependent variables $y_t$ and the predictors $\bF_t$, that is $\left(y_t, \bF_t^\top\right)$, for all $t = 1,\dots, T$.\footnote{Recall that the inclusion of the constant term should be managed via the \code{formula} argument.}
  \item \code{vDelta}: A $d\times 1$ numeric vector representing a grid of $\delta$. Typically we choose the following grid: $\{0.90,~0.91,~\dots,~1.00\}$. By default \code{vDelta = c(0.90, 0.95, 0.99)}.
  \item \code{dAlpha}: A numeric variable representing $\alpha$ in Equation~\ref{eq:3.4}. By default \code{dAlpha = 0.99}.
  \item \code{dBeta}: A numeric variable indicating the forgetting factor for the measurement variance, see Equation \ref{eq:beta} and \citet[p. 132]{prado_west.2010} and \cite{beckmann_schussler.2014}. By default \code{dBeta = 1.0}, \emph{i.e.}, constant observational variance.
  \item \code{vKeep}: A numeric vector of indices representing the predictors that must be always included in the models. The models that do not include the variables declared in \code{vKeep} are automatically discarded. The indices must be consistent with the model description given in \code{formula}. For instance, if the first and fourth variables always have to be included, then we must set vKeep=c(1, 4). Notice that, the intercept (if not removed from \code{formula}) is always in the first position. \code{vKeep} can also be a character vector indicating the names of the predictors if these are consistent with the provided \code{formula}. Furthermore, if \code{vKeep = "KS"} the \qmo Kitchen Sink\qmcsp formulation is adopted, \emph{i.e.}, all the predictors are always included, see, \emph{e.g.}, \cite{Paye.2012}. By default all the combinations are considered, \code{vKeep = NULL}.
  \item \code{bZellnerPrior}: A boolean variable indicating whether the Zellner's prior \citep[see][]{dangl_halling.2012} should be used for the coefficients at time $t=0$. By default \code{bZellnerPrior = FALSE}.
  \item \code{dG}: A numeric variable $(g)$ equal to $100$ by default. If \code{bZellnerPrior = TRUE}, then
      \begin{align}\label{eq:prior}
        p\left(\btheta_{0}^{\left(i\right)}\vert\mathcal{F}_{0}\right) & \sim \mathcal{N}\left(0,gS_{0}^{\left(i\right)}\left(\bF_{1:T}^{\left(i\right)^\top}\bF_{1:T}^{\left(i\right)}\right)^{-1}\right), 
      \end{align}
  where
    \begin{align}
        S_{0}^{\left(i\right)} = & \frac{1}{T-1}\by_{1:T}^\top\left(I_{T}-\bF_{1:T}^{\left(i\right)}\left(\bF_{1:T}^{\left(i\right)^\top}\bF_{1:T}^{\left(i\right)}\right)^{-1}\bF_{1:T}^{\left(i\right)^\top}\right)\by_{1:T},\nonumber
    \end{align}
    and $\by_{1:T}=\left(y_{1},\dots,y_{T}\right)^\top$ and $\bF_{1:T}^{(i)}$ indicating the $T\times p$ design matrix according to model $i$. If \code{bZellnerPrior = FALSE}, it represents the scaling factor for the covariance matrix of the Normal prior for $\btheta^{(i)}_0$, \emph{i.e.}, $\btheta^{(i)}_0\sim N(0,g\times \mathbb{I}),\quad i = 1,\dots,k$, where $\mathbb{I}$ is the identity matrix. We generally recommend practitioners to use the default prior, \emph{i.e.}, \code{bZellnerPrior = FALSE}, especially in the context of quarterly data, where we typically have $200$ to $300$ observations. For longer time--series, results tend to be similar after 100 observations.
  \item \code{bParallelize}: A boolean variable indicating wether to use multiple processors to speed up the computations. By default \code{bParallelize = TRUE}. Since the use of multiple processors is basically effortless for the user, we suggest to not change this value. Furthermore, if the hardware does not permit parallel computations, the program will automatically adapt to run on a single core.
  \item \code{iCores}: An integer indicating the number of cores to use if \code{bParallelize = TRUE}. By default, all but one cores are used. The number of cores is guessed using the
        \code{detectCores()} function from the \pkg{parallel} package. The choice of the number of cores depends on the specific application, namely the length of the time--series $T$ and the number of the predictors $n$. However, as detailed in \cite{chapman_etal.2008}, the level of parallelization of the code should be traded off with the increase in computational time due to threads communications. Consequently, the user can fine tune its application depending on its hardware changing this parameter. Section \ref{sec:computational} reports details about code parallelization.
\end{itemize}

The \code{DMA()} function returns an object of the \textit{formal} class \code{DMA}.\footnote{see, \code{help("class")} and \code{help("DMA-class")}.} This object contains model information and the estimated quantities. It is organized in three slots: \code{model},  \code{Est}, \code{data}. The slot, \code{model}, contains information about the specification used to perform DMA. Examples are: The number of considered models and the computational time in seconds. The slot, \code{Est}, contains the estimated quantities such as: Point forecasts, Predictive likelihood, Posterior inclusion probabilities of the predictors, Filtered estimates\footnote{With the term \qmo filtered estimates\qmcsp we intent estimates at time $t$ conditional on information up to time $t$.} of the regression coefficients, $\btheta_t$ (as in Equation \ref{eq:theta_post}), and so on. Finally, the slot, \code{data}, includes the data passed to the \code{DMA()} function, organised in the vector of responses \code{vY} and a design matrix \code{mF}.

\subsection[Using eDMA]{Using \pkg{eDMA}}
\label{sec:using}

After having installed \pkg{eDMA}, it can be easily loaded using

\begin{CodeChunk}
\begin{CodeInput}
R> library("eDMA")
\end{CodeInput}
\end{CodeChunk}

Thereafter, model estimation can be performed using the R commands reported below.

In order to illustrate how \pkg{eDMA} works in practice, we provide an example based on simulated data. We also provide an application using quarterly inflation data in Section \ref{sec:EmpiricalApplication}. We simulate a time--series of $T=500$ observations from

\begin{equation}\label{eq:dgp}
  y_t = \bF_t^\top\btheta_t + \sqrt{0.1}\varepsilon_t,\quad \varepsilon_t\overset{iid}{\sim}\mathcal{N}\left(0, 1\right).
\end{equation}

The first four elements of $\btheta_t$ vary according to random--walks, whereas the remaining elements in $\btheta_t$ are equal to zero at all time periods. In other words, $\btheta_t = \left(\theta_{1,t}, \theta_{2,t}, \theta_{3,t}, \theta_{4,t}, \theta_{5,t},\theta_{6,t}\right)^\top$ with

\begin{equation}\label{eq:theta_rw}
  \theta_{k,t} = \theta_{k,t-1} + \sqrt{0.01}\eta_{k,t},\quad\eta_{k,t}\overset{iid}{\sim}\mathcal{N}\left(0, 1\right),
\end{equation}

for $k = 1,2,3,4$, and $\eta_{k,t}\indep\eta_{j,t}$, for all $k\neq j$. The last two elements of $\btheta_t$ are equal to zero, that is, $\theta_{5,t}=\theta_{6,t}=0$ for $t=1,\dots, T$. The first element of the $6\times 1$ vector, $\bF_t$, is one, representing the constant term. The remaining elements are generated from a standard Gaussian distribution, \emph{i.e.}, $\bF_t = \left(1.0, x_{2,t}, x_{3,t}, x_{4,t}, x_{5,t}, x_{6,t}\right)^\prime$, where $x_{k,t}\overset{iid}{\sim}\mathcal{N}\left(0, 1\right)$ and $x_{k,t}\indep x_{j,t}$ for all $k\neq j$.
We simulate the data in this way (that is $\theta_{5,t}=\theta_{6,t}=0$) to illustrate that DMA is indeed able to identify the correct variables. In other words, the inclusion probabilities of the last two predictors ought to be zero as they do not impact $y_t$ through $\bF_t$. Conversely, inclusion probabilities of the first four predictors ought to converge to 1.

This data is simulated using the \code{SimulateDLM()} function available in \pkg{eDMA}, details are reported in the \proglang{R} documentation, see \code{help("SimulateDLM")}. We organize the data in a \code{data.frame} named \code{SimData}, which is included in \pkg{eDMA} and can be loaded into the workspace by executing

\begin{CodeChunk}
\begin{CodeInput}
R> data("SimData", package = "eDMA")
\end{CodeInput}
\end{CodeChunk}

DMA is then performed using the function \code{DMA()} as

\begin{CodeChunk}
\begin{CodeInput}
R> Fit <- DMA(y ~ x2 + x3 + x4 + x5 + x6, data = SimData,
          vDelta = seq(0.9, 1.0, 0.01))
\end{CodeInput}
\end{CodeChunk}

Information on the DMA procedure is available by typing:

\begin{CodeChunk}
\begin{CodeInput}
R> Fit
\end{CodeInput}
\begin{CodeOutput}
------------------------------------------
-        Dynamic Model Ageraging         -
------------------------------------------

Model Specification	
T     = 500
n     = 6
d     = 11
Alpha = 0.99
Beta  = 1
Model combinations = 63
Model combinations including averaging over delta = 693
------------------------------------------
Prior : Multivariate Gaussian with mean vector 0
        and covariance matrix equal to: 100 x diag(6)
------------------------------------------
The grid for delta:

Delta =  0.90, 0.91, 0.92, 0.93, 0.94, 0.95,
         0.96, 0.97, 0.98, 0.99, 1.00
------------------------------------------

Elapsed time	: 0.57 secs
\end{CodeOutput}
\end{CodeChunk}

Note, we specify a grid of eleven equally spaced values for $\delta$ ($d=11$) ranging from $0.90$ to $1.00$. Furthermore, since we do not specify any value for \code{bZellnerPrior} and \code{bParallelize}, their default values, \code{bZellnerPrior = FALSE} and \code{bParallelize = TRUE} have been used.

In order to extract the quantities estimated by DMA, the user can relay on the \code{as.data.frame()} method. \code{as.data.frame()} accepts two arguments: (i) An object of the class \code{DMA} and (ii) A character string, \code{which}, indicating the quantity to extract. Possible values for \code{which} are:
\begin{itemize}
  \item \code{"vyhat"}: Point forecasts of DMA, see Equation~\ref{eq:4.5}. \code{"vyhat_DMS"} for point forecast according to DMS.
  \item \code{"mincpmt"}: Posterior inclusion probabilities of the predictors at each point in time, see \cite{koop_korobilis.2012} for more details.
  \item \code{"vsize"}: Expected number of predictors (average size), see \cite{koop_korobilis.2012} and point (ii) at page 7.
  \item \code{"vsize_DMS"}: Number of predictors in the model with the highest posterior model probability, at each point in time, see Equation~\ref{eq:3.2}.
  \item \code{"mtheta"}: Filtered estimates of the regression coefficients for DMA, see Equation~\ref{eq:theta_post}.
  \item \code{"mpmt"}: Posterior probability of the forgetting factors, see Equation~\ref{eq:delta}.
   \item \code{"vdeltahat"}: Posterior weighted average of $\delta$, see point (iv) at page 10 of this paper.
  \item \code{"vLpdfhat"}: Predictive log--likelihood of DMA, see Equation \ref{eq:preddens}.
  \item \code{"vLpdfhat_DMS"}: Predictive log--likelihood of DMS. That is instead of averaging over the individual predictive likelihoods, we select the predictive likelihood of the model combination with the highest posterior probability (\emph{i.e.}, \eqref{eq:3.2}) at each time--period.
  \item \code{"mvdec"}: Individual components of Equation~\ref{eq:varDec}, see point (vi) in page 10 and \cite{dangl_halling.2012} for more details. The function returns a $T\times 5$ matrix whose columns contain the variables.
  \begin{itemize}
    \item \code{vobs}: Observational variance, Obs.
    \item \code{vcoeff}: Variance due to errors in the estimation of the coefficients, Coeff.
    \item \code{vmod}: Variance due to model uncertainty, Mod.
    \item \code{vtvp}: Variance due to uncertainty with respect to the choice of the degrees of time--variation in the regression coefficients, TVP.
       \item \code{vtotal}: Total variance, that is \code{vtotal} = \code{vobs} + \code{vcoeff} + \code{vmod} + \code{vtvp} .
  \end{itemize}
  \item \code{"vhighmp_DMS"}: Highest posterior model probability, \emph{i.e.}, $\underset{i}{\max}~P\left(M_i\vert\mathcal{F}_t\right),\quad t =1,\dots,T$.
  \item \code{"vhighmpTop01_DMS"}: Sum of the $10\%$ highest posterior model probabilities.

\end{itemize}

The additional \code{numeric} argument, \code{iBurnPeriod}, determines the length of the burn--in period, \emph{i.e.}, results before $t$=\code{iBurnPeriod} are discarded. By default, \code{iBurnPeriod = NULL}, meaning that no burn--in period is considered. For instance, in order to extract the posterior inclusion probabilities of the predictors, with a burn--in period of 50 observations, we can easily run the following command

\begin{CodeChunk}
\begin{CodeInput}
R> PostProb <- as.data.frame(Fit, which = "mincpmt", iBurnPeriod = 50)
\end{CodeInput}
\end{CodeChunk}

which returns a $(T-$\code{iBurnPeriod}$)\times 6$ matrix of inclusion probabilities for the predictors at each point in time. Final values of \code{PostProb} are printed as

\begin{CodeChunk}
\begin{CodeInput}
R> round(tail(PostProb), 2)
\end{CodeInput}
\begin{CodeOutput}
       (Intercept) x2 x3 x4   x5   x6
[445,]           1  1  1  1 0.06 0.03
[446,]           1  1  1  1 0.06 0.03
[447,]           1  1  1  1 0.06 0.03
[448,]           1  1  1  1 0.07 0.03
[449,]           1  1  1  1 0.07 0.03
[450,]           1  1  1  1 0.08 0.04
\end{CodeOutput}
\end{CodeChunk}

Furthermore, if the supplied data is a \code{ts}, \code{zoo} or \code{xts} object, the class membership is automatically transferred to the output of the \code{as.data.frame()} method.

The \code{plot()} method is also available for the class \code{DMA}. Specifically, this method prints an interactive menu in the console permitting the user to chose between a series of interesting graphical representation of the estimated quantities. It can be straightforwardly executed running

\begin{CodeChunk}
\begin{CodeInput}
R> plot(Fit)
\end{CodeInput}
\begin{CodeOutput}
Type 1-16 or 0 to exit
 1: Point forecast
 2: Predictive likelihood
 3: Posterior weighted average of delta
 4: Posterior inclusion probabilities of the predictors
 5: Posterior probabilities of the forgetting factors
 6: Filtered estimates of the regression coefficients
 7: Variance decomposition
 8: Observational variance
 9: Variance due to errors in the estimation of the coefficients, theta
10: Variance due to model uncertainty
11: Variance due to uncertainty with respect to the choice of
    the degrees of time-variation in the regression coefficients
12: Expected number of predictors (average size)
13: Number of predictors (highest posterior model probability) (DMS)
14: Highest posterior model probability (DMS)
15: Point forecasts (highest posterior model probability) (DMS)
16: Predictive likelihood (highest posterior model probability) (DMS)
\end{CodeOutput}
\end{CodeChunk}
and selecting the desiderated options. The additional character argument, \code{which}, can be supplied in order to directly plot one particular quantity. Possible values for \code{which} are the same of the \code{as.data.frame()} method. Similar to \code{as.data.frame()}, the additional \code{numeric} argument \code{iBurnPeriod} determines the length of the burn--in period. Typically, it takes around $30$ to $50$ for the model to adapt to the time--series given the prior. Therefore, in almost all applications, the first $30$ to $50$ observations should be discarded.

\begin{figure}[!t]
\centering
\includegraphics[width=1\textwidth]{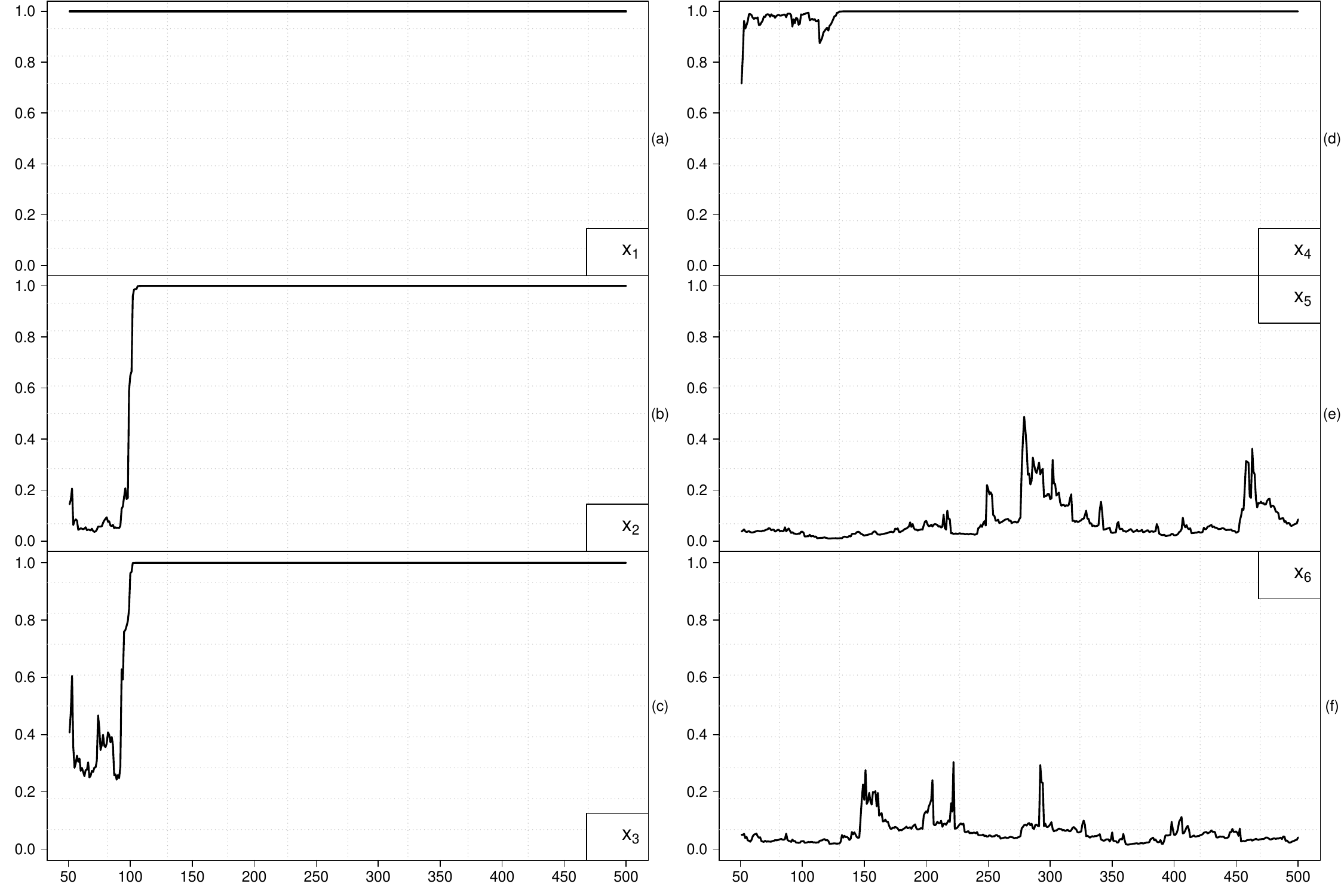}
\caption{Posterior inclusion probabilities of the predictors using simulated data.}
\label{fig:inclusionprob_simulation}
\end{figure}

The code:

\begin{CodeChunk}
\begin{CodeInput}
R> plot(Fit, which = "mincpmt", iBurnPeriod = 50)
\end{CodeInput}
\end{CodeChunk}

plots the inclusion probabilities for the predictors discarding the first 50 observations. The outcome is reported in Figure~\ref{fig:inclusionprob_simulation}. As expected, $x_1$ to $x_4$ quickly converge to $1$ after few observations. Conversely, the inclusion probabilities of the last two predictors with loading factor equal to zero, quickly converge to $0$.

\subsection[Additional methods for the DMA class]{Additional methods for the \code{DMA} class}
\label{sec:methods}

The \code{DMA} class comes with several methods for extracting and representing estimated quantities. The \code{plot()}, \code{as.data.frame()} and \code{show()} methods have been previously introduced, additional methods are: \code{summary()}, \code{coef()}, \code{residuals()}, \code{inclusion.prob()}, and \code{pred.like()}.

For instance, the \code{summary} method prints a summary of the estimated model directly in the console. The code:

\begin{CodeChunk}
\begin{CodeInput}
R> summary(Fit, iBurnPeriod = 50)
\end{CodeInput}
\end{CodeChunk}

produces the output:

\begin{CodeChunk}
\begin{CodeOutput}
Call:
 DMA(formula =  y ~ x2 + x3 + x4 + x5 + x6 )

Residuals:
    Min      1Q  Median      3Q     Max
-2.0445 -0.3844  0.0414  0.4398  2.3759

Coefficients:
            E[theta_t] SD[theta_t] E[P(theta_t)] SD[P(theta_t)]
(Intercept)       0.51        0.68          1.00           0.00
x2               -0.64        0.65          0.90           0.29
x3                2.10        1.74          0.92           0.23
x4               -1.43        1.02          0.99           0.03
x5                0.01        0.03          0.07           0.07
x6                0.00        0.01          0.06           0.04

Variance contribution (in percentage points):
  vobs vcoeff   vmod   vtvp
 64.12  34.24   1.50   0.15

Top 10% included regressors:	(Intercept)

Forecast Performance:
                             DMA      DMS
MSE                        0.489    0.483
MAD                        0.539    0.532
Log-predictive Likehood -463.820 -463.076
\end{CodeOutput}
\end{CodeChunk}

where the quantities, \code{E[theta_t], SD[theta_t], E[P(theta_t)]} and \code{SD[P(theta_t)]} represent the means and standard deviations across the time dimension of the filtered estimates of $\btheta_t$, and the inclusion probabilities after burn-in.

The last part of the summary, (\code{Forecast Performance}), prints the output of the \code{BacktestDMA()} function implemented in \pkg{eDMA}. \code{BacktestDMA()} accepts a \code{DMA} object and returns a \code{matrix} with out--of--sample mean squared error (MSE), mean absolute deviation (MAD) and log--predictive likelihood, computed according to DMA and DMS, see \code{help("BacktestDMA")}.

The additional methods: \code{coef()}, \code{residuals()}, \code{inclusion.prob()}, and \code{pred.like()} are wrapper to the \code{as.data.frame()} method and focus on particular estimated quantities, for instance:

\begin{itemize}
  \item[-] \code{coef()}: Returns a $T\times n$ \code{matrix} with the filtered regressor coefficients, $\btheta_t,\quad t = 1,\dots,T$.
  \item[-] \code{residuals()}: Extract the residuals of the model, \emph{i.e.}, $y_t - \hat{y_t},\quad t = 1,\dots, T$. The additional \code{boolean} argument \code{standardize} controls if the standardize residuals should be returned. By default \code{standardize = FALSE}. The additional \code{character} argument, \code{type}, permits to choose between residuals evaluated using DMA (\code{"DMA"}) or DMS (\code{"DMS"}). By default \code{Type = "DMA"}.
  \item[-] \code{inclusion.prob()}: Extract the inclusion probabilities of the predictors. Analogous to \code{as.data.frame(object, which = "mincpmt", iBurnPeriod)}.
  \item[-] \code{pred.like()}: Extract the predictive log--likelihood series. The additional argument \code{type} permits to choose between predictive likelihoods evaluated using DMA and DMS. By default \code{Type = "DMA"}. Similar to the above variables, \code{pred.like()} accepts \code{iBurnPeriod}.
  \item[-] \code{getLastForecast}: If we extend the time--series of the dependent variable of length $T$ (\emph{i.e.}, observations that we actually observe till time $T$) with an \code{NA}, resulting in a series of length $T+1$, then the \code{DMA()} function computes the point forecast and the associated variance decomposition for the future observation at time $T+1$, see Appendix \ref{sec:app2} for further details. In this case, the \code{getLastForecast} can be used to extract the \qmo true\qmcsp out--of--sample\footnote{We use the term \qmo true\qmcsp out---of--sample to distinguish from the case of \qmo pseudo\qmcsp out--of--sample which consists to the usual recursive out--of--sample forecasts, where one compares the forecasts with the actual observed values.} forecast at time $T+1$.
\end{itemize}

\section[Computational challenges]{Computational challenges}
\label{sec:computational}
Although estimation of DMA does not require resorting to simulation, in many economic applications, performing DMA can become computationally cumbersome.
As it can be seen from the set of recursions from
the Section \ref{sec:ModifiedDMA}, DMA consists of a large number of model combinations, where a lot of the quantities must be saved for subsequent analysis. Therefore, in many cases, DMA tends to occupy a large chunk of Random--Access Memory (RAM). Often on a standard PC, the system basically runs out of memory due to the large number of combinations and the amount of information that must be saved. Therefore, it limits the use of DMA to middle--sized data--sets. For instance, in their seminal paper, \cite{koop_korobilis.2012} use DMA to forecast quarterly inflation. Thus, $y_{t}$ in Equation~\ref{eq:2.1} is the percentage changes in the quarterly U.S.
GDP price deflator and $\bF_{t}$ consists of $14$ exogenous predictors and three lags of $y_{t}$ for a total of $17$ variables. However, handling $2^{17}$
combinations even in the context of quarterly data, which at most consists of around 300 observations, reveals to be cumbersome in their programming framework. Therefore, \cite{koop_korobilis.2012}
choose to include three lags of inflation in all model combinations and thus reduce the model space to $2^{14}$ model combinations. Furthermore, they do not consider a grid for different values of $\delta$, which would result in $2^{14}\times d$ combinations, making inference even more challenging.

\begin{figure}
\centering
\includegraphics[width=1\textwidth]{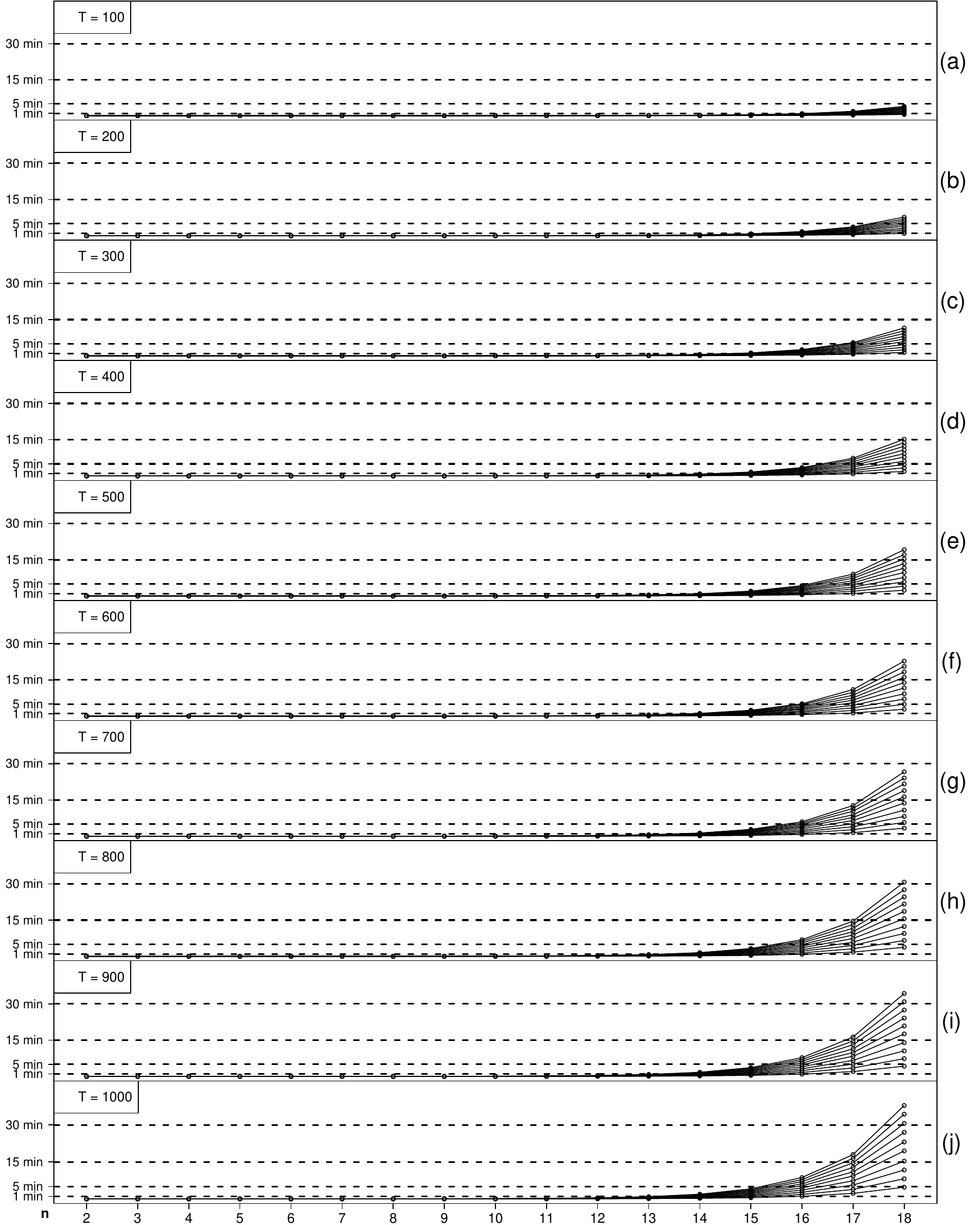}
\caption{Computational time for \code{DMA()} using simulated data. Each panel represents computation time in minutes for \code{DMA()} using different sample sizes, $T$, number of predictors, $n$, and values of $d$, the number of points in the grid of $\delta$. The values for $d$ range between 2 and 10, the solid line at the bottom of each subfigure is for $d=2$, the one immediately above is $d=3$ and so on until the last which is for $d=10$. Computations are performed on a standard Intel Core i7--4790 processor with 8 threads and 8 GB of RAM with Ubuntu 12.04 server edition.}
\label{fig:comptime}
\end{figure}

We can argue that DMA can impose a substantial challenge for the practitioner when dealing with a large number of predictors and high number of observations,
namely that, besides dealing with the task of transforming mathematical
equations from paper to codes, handling data and estimation
issues, practitioners also has to overcome \qmo technical/computer science related\qmcsp challenges
such as how to deal with extensive memory consumption and how to use multiple cores instead of a single core to speed up computation time.
Although one can always improve the computational procedure by \qmo coding smarter\qmcsp or discovering ways to optimize memory allocation,
it seems unreasonable to expect that practitioners in economics
should have extensive knowledge of computer science concepts
such as those stated above.

In this paper, we provide practical solutions to these problems. First, reduction in computation time is implemented by writing all the code in \proglang{C++} using the \code{armadillo} library of \cite{sanderson.2010}.
Second, we  exploit multiple processors through the \code{OpenMP} API whenever the hardware is suited for that.
The combination of \proglang{C++} routines and parallel processing permits to dramatically speed up the computations over the same code written in plain \proglang{R}.

In order to provide an intuitive example of the advantages of our package, we report a comparison between our code and the available \pkg{dma} package of \cite{dma}. Note that, the \pkg{dma} package is entirety written in plain \proglang{R} and cannot be run in parallel, consequently, even if the algorithm we implement is slightly different from those of \pkg{dma} \citep[recall that we follow the implementation of][]{dangl_halling.2012}, improvement in computational time should be principally attributed to the two aforementioned reasons.

For this experiment, since the \pkg{dma} package cannot operate over a grid value of $\delta$, we fix $\delta$ at $0.95$. We simulate $T=\{100,500,1000\}$ observations from a DLM with $n = \{4,6,8,10,12,14,16\}$ predictors and evaluate the differences in the computational time of the \code{dma()} function in the \pkg{dma} package and the \code{DMA()} function in the presented \pkg{eDMA} package. The experiment is performed on a standard Intel Core i7--4790 processor with 8 threads and Ubuntu 12.04 server edition.

Table~\ref{tab:RafComparison} reports the ratio of the CPU time for different values of $T$ and $n$ between \code{dma()} and \code{DMA()}. As one can note, the decrease in computational time in favor of our package is huge. For example, in the case $T = 500$ and $n = 16$, \code{dma()} takes 37.57 minutes while \code{DMA()} only 1.48.\footnote{Also note that this cannot be considered as a one to one comparison because \code{DMA()} performs additional operations (such as DMS and variance decomposition) which are not considered by \code{dma()}. Furthermore, the time for the construction of the power set of all possible model combinations has not been included for \code{dma()}.}
It is also worth stressing that, the benefit of using \pkg{eDMA} does not only concern the possibility of running moderately large applications in a reasonable time using a commercial hardware, but also enables practitioners to run application with a large number of exogenous variables. To give an idea of the computational time a \pkg{eDMA} user faces, we report a second simulation study. We simulate from a DLM with $T = \{100,200,\dots,900,1000\}$, $n = \{2,3,\dots,18\}$ and run \code{DMA()} using a grid of values for $\delta$ between $0.9$ and $1.0$ with different spaces $d$, namely $d=\{2,3,\dots,10\}$. Figure~\ref{fig:comptime} displays the computational time in minutes for all the combinations of $T,n,d$. The lines reported in each subfigure represent the computational time for a specific choice of $d$. The line at the bottom of each subfigure is for $d=2$,\footnote{In this case $\delta$ can take values $\delta=0.9$ and $\delta = 1.0$.} the one immediately above is for $d=3$ and so on until $d=10$. From the Figure, we can see that, when $T\le400$, even for $n=18$ and $d=10$, the computational time is less then 15 minutes. Such sample sizes are relatively common in economic applications. When $T$ increases, computational time increases linearly. For example, when $T=800, n = 18$ and $d=10$, computational time is 30 minutes, which is the double of the same case with $T=400$.

The other relevant problem with DMA is the RAM usage. Specifically, if we want to store the quantities defined in Equations \ref{eq:2.1} and \ref{eq:3.3}, we need to define two arrays of dimension $T\times d\times k$. These kind of objects are not present in the \pkg{eDMA} package since we rely on the markovian nature of the model clearly evident from Equation~\ref{eq:2.1}. In this respect, ,we keep track of the quantities coming from Equation~\ref{eq:3.3} and $p\left(y_{t}\vert M_{i},\delta_{j},\mathcal{F}_{t-1}\right)$ only for two consecutive periods during the loop over $T$.\footnote{Differently, in the \pkg{dma} package a full $T\times k$ matrix is stored.} RAM usage is still efficiently performed in the \pkg{eDMA} package. Indeed, the computer where we run all our simulations has only $8$GB of RAM. A formal analysis of RAM usage with the \pkg{eDMA} package is hard to implement given that RAM profiling for \proglang{C++} functions wrapped in \proglang{R} cannot be easily performed\footnote{This is the case also for contributed packages such as \pkg{profvis} of \cite{profvis}.}. However, we find that \pkg{eDMA} on a Windows 10 based system equipped with $16$GB of RAM fixing $T = 300$ is able to handle 4'194'304 model combinations while, for example, \pkg{dma} only 2'097'157, \emph{i.e.}, half of \pkg{eDMA}.

\begin{table}[!t]
\centering
\setlength\tabcolsep{17pt}
\begin{tabular}{lccccccc}
\toprule
$T/n$ & 4 & 6 & 8 & 10 & 12 & 14 & 16\\
\cmidrule(lr){1-1}\cmidrule(lr){2-8}
100 & 10.9 & 92.6 & 133.2 & 88.4 & 69.4 & 70.5 & 58.4 \\
500 & 37.5 & 29.4 & 31.5 & 30.7 & 25.7 & 26.5 & 25.4 \\
1000 & 13.0 & 15.0 & 13.8 & 12.9 & 12.7 & 13.5 & 13.8 \\
\bottomrule
\end{tabular}
\caption{\footnotesize{Ratio of computation time between the \code{dma()} function from the \pkg{dma} package of \cite{dma} and the \code{DMA()} function of the \pkg{eDMA} package using different values of $T$ and $n$.}}
\label{tab:RafComparison}
\end{table}

\section[A DMA example: Inflation data]{A DMA example: Inflation data}
\label{sec:EmpiricalApplication}

We use a time--series of quarterly U.S. inflation rate with exogenous predictors for illustration and then step by step show how to obtain
posterior output. The example can be thought
of as a typical assignment for a researcher at a central bank who
is interested in forecasting inflation several--quarters ahead and understand the relationship between inflation, business cycles and perform variance decomposition.

\subsection[Data]{Data}

We rely on the data--set of \cite{groen_etal.2013}.\footnote{The data is downloadable from \url{http://www.tandfonline.com/doi/suppl/10.1080/07350015.2012.727718}.}
As a measure of inflation, $y_{t}$, we consider quarterly
log--changes in the Gross Domestic Product implicit price deflator (GDPDEF) ranging from 1960q1 to 2011q2. The number
of exogenous predictors are fifteen. This number is in accordance
with typical \qmo real--world\qmcsp applications, see also \cite{dangl_halling.2012} and \cite{koop_korobilis.2012}.

We start by loading the \pkg{eDMA} package and the data--set by typing:

\begin{CodeChunk}
\begin{CodeInput}
R> library("eDMA")
R> data("USData", package = "eDMA")
\end{CodeInput}
\end{CodeChunk}

The predictors are: Real GDP in volume terms (ROUTP), real
durable personal consumption expenditures in volume terms (RCONS), real residential investment in
volume terms (RINVR), the import deflator (PIMP), the unemployment
ratio (UNEMP), non--farm payrolls data on employment (NFPR), housing
starts (HSTS), the real spot price of oil (OIL), the real food commodities
price index (FOOD) the real raw material commodities price index
(RAW), and the M2 monetary aggregate (M2), which can reflect information
on the current stance of monetary policy and liquidity in the economy
as well as spending in households. In addition, we also use data on
the term structure of interest rates approximated by means of:
The level factor (YL), the slope factor (TS) and curvature factor
(CS). Finally, we proxy inflation expectations through the one--year
ahead inflation expectations that come from the Reuters/Michigan Survey
of Consumers (MS). We include the data (the GDPDEF series along with the fifteen predictors) in the \pkg{eDMA} package as a \code{xts} object of dimension $206\times 16$ named \code{USData}.
A glance of GDPDEF series and the first five predictors is obtained by typing:

\begin{CodeChunk}
\begin{CodeInput}
R> head(round(USData[,1:6], 2))
\end{CodeInput}
\begin{CodeOutput}
           GDPDEF ROUTP RCONS RINVR  PIMP UNEMP
1960-01-01  -1.14  1.66  0.62  0.55 -0.48 -0.56
1960-04-01  -0.77 -1.39  0.33 -1.84 -0.37 -0.49
1960-07-01  -0.71 -0.68 -0.71 -0.69 -0.16 -0.30
1960-10-01  -0.76 -2.32 -1.27 -0.07 -0.47  0.16
1961-01-01  -1.27 -0.19 -2.25  0.04 -0.32  0.49
1961-04-01  -1.16  1.24  0.23  0.03 -0.40  0.62
\end{CodeOutput}
\end{CodeChunk}

For most series, we follow \cite{groen_etal.2013} and use the percentage change of the original series
in order to remove possible stochastic and deterministic trends. Exceptions
are HSTS, for which we use the logarithm of the respective levels,
as well as UNEMP, YL, TS, CS and MS, where we use the \qmo raw\qmcsp levels,
see \cite{groen_etal.2013} for more details. Finally, since inflation is very persistence, besides these $15$ predictors, we follow \cite{groen_etal.2013} and also include four inflation lags,
$y_{t-1},\dots,y_{t-4}$, as predictors. In \pkg{eDMA}, we implement the function, \code{Lag()}, which allows us to lag variables delivered in the form of vector or matrices. For instance, to lag the \code{numeric} vector \code{X} of length $T$ by one period, we simply run

\begin{CodeChunk}
\begin{CodeInput}
R> Lag(X, 1)
\end{CodeInput}
\end{CodeChunk}

which returns a \code{numeric} vector of length $T$ containing the lagged values of \code{X}. Values that are not available are replaced by \code{NA}.

\subsection[Model estimation]{Model estimation}

We have a total of $2^{19} = 524288$ model combinations.\footnote{Models which do not include the constant term are not considered. Note that, when \code{vKeep = NULL}, the number of models is $2^n - 1$, however, when \code{vKeep != NULL}, the number of models is $2^b - 1$, where \code{b = n - length(vKeep)}.} Furthermore, we let $\delta=\left\{ 0.9,0.91,...,1\right\} $ such
that we have a total of $\left(2^{19}\right)\cdot11 = 5767168$ combinations. We
set $\beta=0.96$, a value we generally suggest in the context of working with quarterly data, $\alpha=0.99$, $g=100$, $p\left(M_{s}\mid\mathcal{F}_{0}\right)=1/\left(d\cdot k\right),\quad s = 1,\dots,d\cdot k$,
such that initially, all models are equally likely. We then update these model
probabilities as new information arrives. As previously mentioned,
we include a constant term in all models, see also \cite{groen_etal.2013}.

In order to perform DMA using the \code{DMA()} function, we write\footnote{Note that this command can be computational expensive for non--OpenMP ready systems.}:

\begin{CodeChunk}
\begin{CodeInput}
R> Fit <- DMA(GDPDEF ~  Lag(GDPDEF, 1) + Lag(GDPDEF, 2) +
                        Lag(GDPDEF, 3) + Lag(GDPDEF, 4) +
                        Lag(ROUTP, 1)  + Lag(RCONS, 1)  +
                        Lag(RINVR, 1)  + Lag(PIMP, 1)   +
                        Lag(UNEMP, 1)  + Lag(NFPR, 1)   +
                        Lag(HSTS, 1)   + Lag(M2, 1)     +
                        Lag(OIL, 1)    + Lag(RAW, 1)    +
                        Lag(FOOD, 1)   + Lag(YL, 1)     +
                        Lag(TS, 1)     + Lag(CS, 1)     +
                        Lag(MS, 1), data = USData,
                        vDelta = seq(0.90, 1.00, 0.01), vKeep = 1,
                        dBeta = 0.96, dAlpha = 0.99)
\end{CodeInput}
\end{CodeChunk}

We suggest using the non--informative prior,
\code{bZellnerPrior = FALSE}, which is the default. This, way the regression coefficients are centered at 0 with a flat prior and adapt quickly in the averaging process as new information arrives. More details on the model can be made available by typing \code{Fit}

\begin{CodeChunk}
\begin{CodeInput}
R> Fit
\end{CodeInput}
\begin{CodeOutput}
------------------------------------------
-        Dynamic Model Ageraging         -
------------------------------------------

Model Specification	
T     = 202
n     = 20
d     = 11
Alpha = 0.99
Beta  = 0.96
Model combinations = 524288
Model combinations including averaging over delta = 5767168
------------------------------------------
Prior : Multivariate Gaussian with mean vector 0
        and covariance matrix equal to: 100 x diag(20)

Variables always included : (Intercept)
------------------------------------------
The grid for delta:

Delta =  0.90, 0.91, 0.92, 0.93, 0.94, 0.95,
         0.96, 0.97, 0.98, 0.99, 1.00
------------------------------------------

Elapsed time : 1429.13 secs
\end{CodeOutput}
\end{CodeChunk}

As it can be seen, the total estimation time of our DMA when working with more than 5'700'000 model combinations at each time--period is 1429.13 seconds corresponding to around
23.8 minutes on an Intel Core i7-3630QM processor. A complete summary of the estimation is available as:

\begin{CodeChunk}
\begin{CodeInput}
R> summary(Fit, iBurnPeriod = 32)
\end{CodeInput}
\begin{CodeOutput}
Call:
 DMA(formula =  Lag(GDPDEF, 1) + Lag(GDPDEF, 2) +
                Lag(GDPDEF, 3) + Lag(GDPDEF, 4) +
                Lag(ROUTP, 1)  + Lag(RCONS, 1)  +
                Lag(RINVR, 1)  + Lag(PIMP, 1)   +
                Lag(UNEMP, 1)  + Lag(NFPR, 1)   +
                Lag(HSTS, 1)   + Lag(M2, 1)     +
                Lag(OIL, 1)    + Lag(RAW, 1)    +
                Lag(FOOD, 1)   + Lag(YL, 1)     +
                Lag(TS, 1)     + Lag(CS, 1)     +
                Lag(MS, 1) )

Residuals:
    Min      1Q  Median      3Q     Max
-1.3948 -0.3169 -0.0073  0.2309  1.6503

Coefficients:
               E[theta_t] SD[theta_t] E[P(theta_t)] SD[P(theta_t)]
(Intercept)          0.08        0.16          1.00           0.00
Lag(GDPDEF, 1)       0.43        0.17          0.84           0.29
Lag(GDPDEF, 2)       0.03        0.02          0.20           0.13
Lag(GDPDEF, 3)       0.10        0.08          0.38           0.25
Lag(GDPDEF, 4)       0.10        0.05          0.42           0.21
Lag(ROUTP, 1)        0.00        0.01          0.13           0.09
Lag(RCONS, 1)        0.00        0.00          0.12           0.07
Lag(RINVR, 1)        0.01        0.02          0.13           0.07
Lag(PIMP, 1)         0.19        0.08          0.77           0.29
Lag(UNEMP, 1)       -0.03        0.09          0.12           0.10
Lag(NFPR, 1)         0.02        0.02          0.20           0.16
Lag(HSTS, 1)         0.02        0.02          0.16           0.08
Lag(M2, 1)           0.01        0.01          0.16           0.08
Lag(OIL, 1)         -0.02        0.05          0.22           0.23
Lag(RAW, 1)          0.00        0.01          0.11           0.07
Lag(FOOD, 1)         0.01        0.01          0.17           0.12
Lag(YL, 1)           0.20        0.34          0.25           0.29
Lag(TS, 1)           0.00        0.01          0.11           0.05
Lag(CS, 1)          -0.02        0.04          0.14           0.07
Lag(MS, 1)           0.02        0.03          0.15           0.07

Variance contribution (in percentage points):
  vobs vcoeff   vmod   vtvp
 65.70  13.21  19.93   1.16

Top 10% included regressors:	(Intercept), Lag(GDPDEF, 1)

Forecast Performance:
                            DMA      DMS
MSE                       0.226    0.278
MAD                       0.355    0.386
Log-predictive Likehood -98.490 -121.752
\end{CodeOutput}
\end{CodeChunk}

Note that, we set burn--in to 32 (\code{iBurnPeriod = 32}) such that the start of the evaluation period corresponds to 1969q1, see also \cite{koop_korobilis.2012}. Below, we go into more details with regards to how to use the output from the estimation procedure.

\subsection[Using the output from eDMA]{Using the output from \pkg{eDMA}}

The output can be divided into two main parts: (a): Full--sample, (b): Out--of--sample analysis. With regards to (a), the most
interesting quantities are:~\code{mincpmt},~\code{vsize}, ~\code{mtheta},~
\code{vdeltahat}, and \code{mvdec}, see Section \ref{sec:ThePackage}.

For instance, the inclusion probabilities of the predictors for the last part of the sample can be
printed by:

\begin{CodeChunk}
\begin{CodeInput}
R> InclusionProb <- inclusion.prob(Fit, iBurnPeriod = 32)
R> tail(round(InclusionProb[, 1:4], 2))
\end{CodeInput}
\begin{CodeOutput}
            (Intercept) Lag(GDPDEF, 1) Lag(GDPDEF, 2) Lag(GDPDEF, 3)
2010-01-01           1           0.99           0.48           0.71
2010-04-01           1           0.99           0.49           0.72
2010-07-01           1           0.99           0.51           0.73
2010-10-01           1           0.99           0.51           0.73
2011-01-01           1           0.99           0.51           0.73
2011-04-01           1           0.99           0.51           0.73
\end{CodeOutput}
\end{CodeChunk}

\begin{figure}[!t]
\centering
\includegraphics[width=1\textwidth]{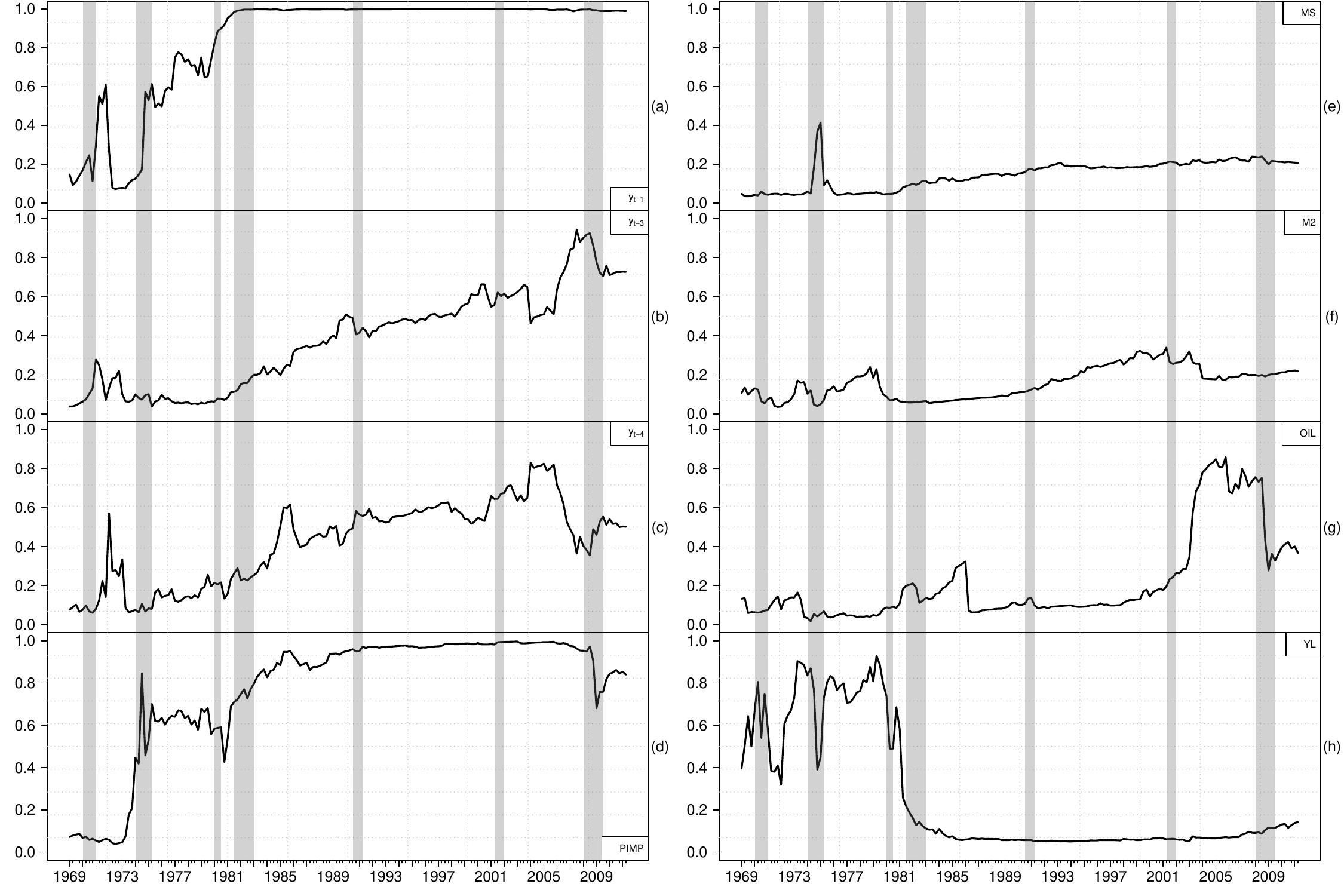}
\caption{Posterior inclusion probabilities for the most important predictors of DMA. Panels (a), (b) and (c): First, third and fourth lags of inflation. Panel (d): Import deflator (PIMP). Panel (e): Inflation expectations (MS). Panel (f): M2 monetary aggregate (M2). Panel (g): Real spot price of oil (OIL). Panel (h): Level factor of the term structure (YL). We refer the reader to \cite{groen_etal.2013} for more details regarding the variables. The gray vertical bars indicate business cycle peaks, \emph{i.e.}, the point at which an economic expansion transitions to a recession, based on National Bureau of Economic Research (NBER) business cycle dating.}
\label{fig:incprob}
\end{figure}

The above matrix shows the inclusion probabilities of: The constant and
$y_{t-1},...,y_{t-3}$, from $2010q1$ to $2011q2$. Notice that, the inclusion probabilities of the
constant term, \code{(Intercept)}, are always equal to $1$ as every model contains this
term (since we set \code{vKeep = 1}), see (iii) in page 10 of this paper. The interested reader can examine these estimates more carefully.

\begin{figure}[!t]
\centering
\includegraphics[width=1\textwidth]{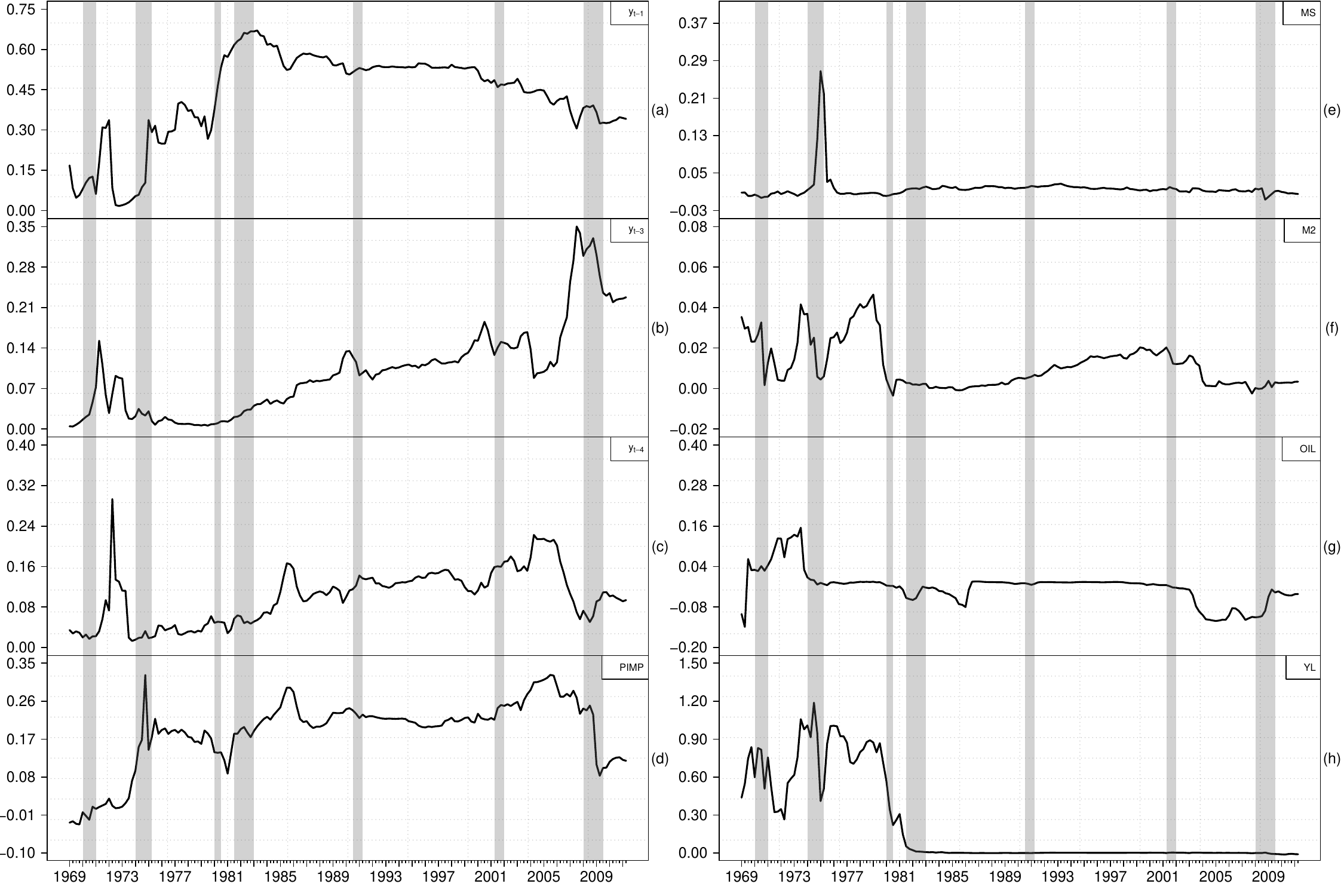}
\caption{Filtered estimates of the regression coefficients for the most important predictors of DMA. Panels (a), (b) and (c): First, third and fourth lags of inflation. Panel (d): Import deflator (PIMP). Panel (e): Inflation expectations (MS). Panel (f): M2 monetary aggregate (M2). Panel (g): Real spot price of oil (OIL). Panel (h): Level factor for the terms structure (YL). We refer the reader to \cite{groen_etal.2013} for more details regarding the variables. The gray vertical bars indicate business cycle peaks, \emph{i.e.}, the point at which an economic expansion transitions to a recession, based on National Bureau of Economic Research (NBER) business cycle dating.}
\label{fig:beta}
\end{figure}

In Figure~\ref{fig:incprob}, we report the inclusion probabilities for the more important predictors. To be precise, any
predictor where the inclusion probabilities are never above 0.2 is excluded.
In these plots, we also make evident NBER recorded recessions
(shaded gray bars). Overall, we observe a good amount of time--variation
these plots. The lags of inflation, except for $y_{t-2}$ all seem
important. The import deflator (PIMP) also receives high posterior
probability throughout the sample. Inflation expectation (MS) and
M2 receive higher probabilities towards the end of the sample. Real spot price of oil (OIL) receives high inclusion probabilities
during the post Great Moderation era, whereas we observe the opposite
trend for YL. In addition to the inclusion probabilities, we also report filtered
estimates of the regression coefficients for these predictors in Figure~\ref{fig:beta}. These quantities are extracted from \code{Fit} simply using

\begin{CodeChunk}
\begin{CodeInput}
R> mTheta <- coef(Fit, iBurnPeriod = 32)
\end{CodeInput}
\end{CodeChunk}

Besides these variables, the output from DMA can be used to analyze:

\begin{figure}[!t]
\centering
\includegraphics[width=1\textwidth]{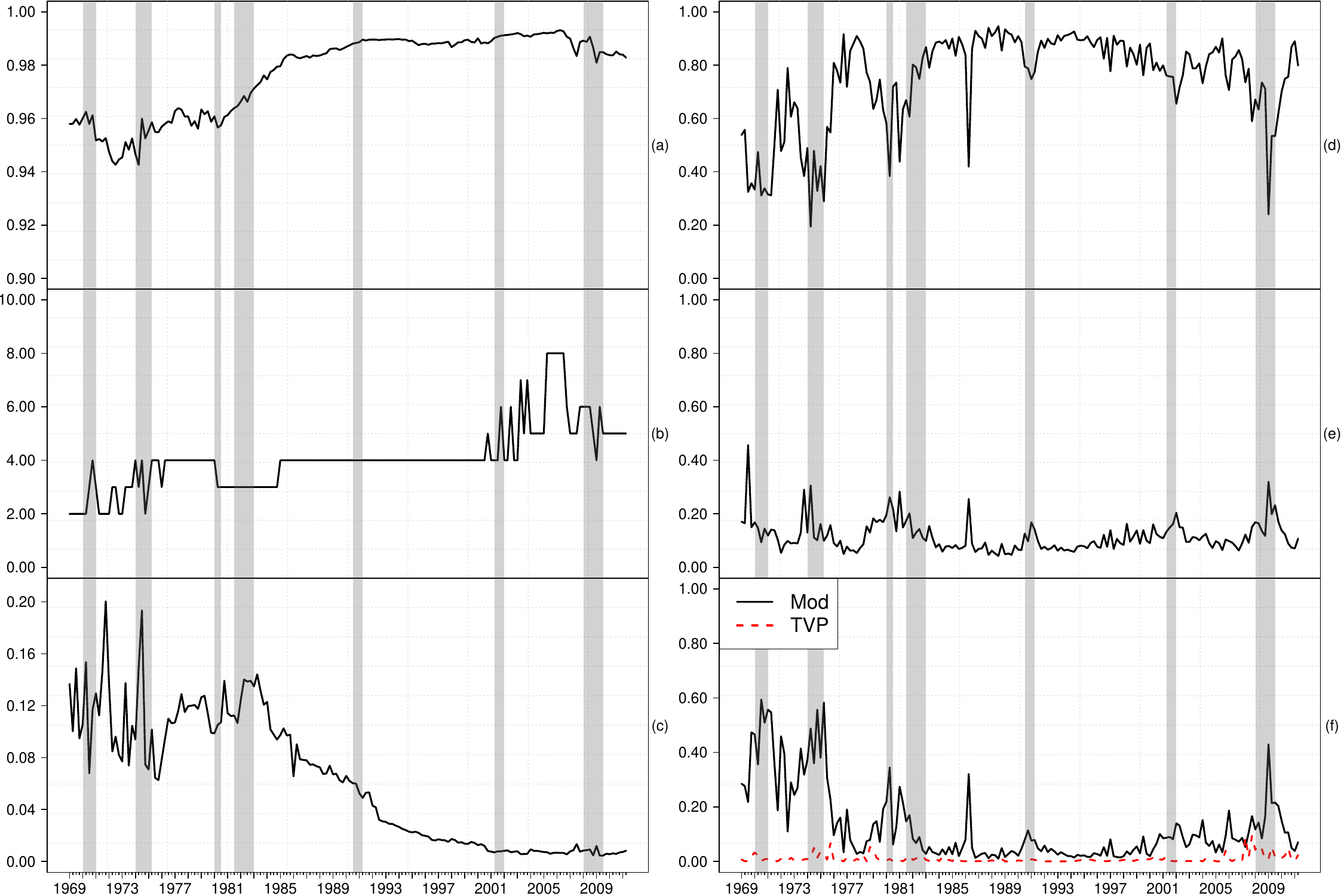}
\caption{Posterior output for DMA. Panel (a): Posterior weighted average estimate of $\delta$. Panel (b):  Number of predictors for the model with the highest posterior probability. Panel (c):  Sum of top $10\%$ inclusion probabilities. Panel (d): Observational variance. Panel (e): Variance due to errors in the estimation of the coefficients. Panel (f) Variance due to model uncertainty (Mod, \textit{solid}) and variance due to uncertainty with respect to the choice of the degrees of time-variation in the regression coefficients (TVP, \textit{red-dotted}). The gray vertical bars indicate business cycle peaks, \emph{i.e.}, the point at which an economic expansion transitions to a recession, based on National Bureau of Economic Research (NBER) business cycle dating.}
\label{fig:combined}
\end{figure}

The magnitude of time--variation in the regression coefficients, \code{"vdeltahat"},
which is the posterior weighted average of $\delta$ at each point in
time. We report this estimate in panel (a) of Figure~\ref{fig:combined}. The analogous plot in \proglang{R} can be obtained using:

\begin{CodeChunk}
\begin{CodeInput}
R> plot(Fit, which = "vdeltahat", iBurnPeriod = 32)
\end{CodeInput}
\end{CodeChunk}

There is a very intuitive relationship between $\delta$ and the business cycles.
Typically, $\delta$ falls at the onset of recessions, which fares well with
the notion that relatively larger shocks hit $\btheta_t$ in these periods. Thereafter, $\delta$ tends to rise again. Conversely, $\delta$ remains high
and close to $1$ during the Great Moderation, which again fares well with the notion of relatively minor variation in the regression coefficients in expansion periods. We can also use \code{as.data.frame()} to extract the posterior probability of each value of $\delta$ and print them using:

\begin{CodeChunk}
\begin{CodeInput}
R> InclusionProbDelta <- as.data.frame(Fit, which = "mpmt", iBurnPeriod = 32)
R> round(tail(InclusionProbDelta), 2)
\end{CodeInput}
\begin{CodeOutput}
           0.9 0.91 0.92 0.93 0.94 0.95 0.96 0.97 0.98 0.99    1
2010-01-01   0    0 0.01 0.01 0.01 0.02 0.05 0.10 0.21 0.31 0.27
2010-04-01   0    0 0.01 0.01 0.01 0.03 0.05 0.10 0.21 0.31 0.26
2010-07-01   0    0 0.00 0.00 0.01 0.02 0.04 0.10 0.22 0.33 0.27
2010-10-01   0    0 0.00 0.00 0.01 0.02 0.05 0.12 0.23 0.32 0.24
2011-01-01   0    0 0.00 0.00 0.01 0.02 0.05 0.12 0.23 0.31 0.24
2011-04-01   0    0 0.00 0.01 0.01 0.02 0.06 0.13 0.25 0.31 0.21
\end{CodeOutput}
\end{CodeChunk}

where the column names are the values of $\delta$.

In panel (b) of Figure~\ref{fig:combined}, we report the number of predictors contained in the model with the
highest posterior probability, $p\left(M_i\vert \mathcal{F}_t\right)$, at each point in time. This can be achieved by:

\begin{CodeChunk}
\begin{CodeInput}
R> plot(Fit, which = "vsize_DMS", iBurnPeriod = 32)
\end{CodeInput}
\end{CodeChunk}

We can also plot the expected number of predictors replacing \code{which = "vsize_DMS"} by \code{which = "vsize"}. An interesting result from
panel (b) is that, although we have $19$ predictors, at each point
in time the best model contains only a few predictors.  We can also use posterior model probabilities to obtain an idea of how important is model averaging. In panel (c), we report the sum of
the posterior inclusion probabilities for the $10\%$ of models (\code{which = "vhighmpTop01_DMS"}). If this number is high, then it means that
relatively few model combinations dominate, and thus obtain relatively high posterior probabilities. Conversely,
if this number is low, then no individual (or group of) model combinations receive
high probabilities, which provides evidence in favor of averaging over predictors.

Finally, in panels (d), (e) and (f) of Figure~\ref{fig:combined}, we report the variance decomposition
analysis (\code{which = "mvdec"}). Evidently, the dominant source of uncertainty is the observational
variance. This is not surprising as random fluctuation are expected to dominate uncertainty. Furthermore, uncertainty regarding the degree of time--variation in the regression (TVP) is relatively lower. However, this is understandable as posterior probabilities of $\delta$ (see above) favor $\delta=0.98$, $0.99$ and $1$.

\begin{table}[h]
\setlength\tabcolsep{3.0pt}{}%
\begin{tabular}{cl}
\hline
Model & Description\\
\hline\\
$\mathcal{M}_{0}$ & \parbox{14cm}{Plain AR(4) model: The constant term and $y_{t-1},\dots,y_{t-4}$ are always included. We set $\alpha=1$, $\delta=1$ and $\beta = 1$.}\\[15pt]

$\mathcal{M}_{1}$ & \parbox{14cm}{Time-varying AR(4) model: The constant term and $y_{t-1},...,y_{t-4}$ are always included. We set $\alpha=0.99$, $\beta = 0.96$ and average over $\delta_{1},\dots,\delta_{d}$.}\\[15pt]

$\mathcal{M}_{2}$ & \parbox{14cm}{DMA using $y_{t-1},\dots,y_{t-4}$: The constant term is always included. We set $\alpha=0.99$, $\beta = 0.96$ and average over the combinations of $y_{t-1},...,y_{t-4}$ and $\delta_{1},\dots,\delta_{d}$.}\\[15pt]

$\mathcal{M}_{3}$ & \parbox{14cm}{DMA using $y_{t-1},\dots,y_{t-4}$ and the exogenous predictors: The constant term is always included. We set $\alpha=0.99$, $\beta = 0.96$ and average over the combinations of predictors as well as $\delta_{1},\dots,\delta_{d}$.}\\[18pt]

$\mathcal{M}_{4}$ & \parbox{14cm}{DMS using $y_{t-1},\dots,y_{t-4}$ and the exogenous predictors: The constant term is always included. We set $\alpha=0.99$, $\beta = 0.96$ and select the model with the highest posterior probability at each $t$ and use it to forecasts.}\\[18pt]

$\mathcal{M}_{5}$ & BMA: DMA with $\alpha=1$, $\delta=1$ and $\beta = 1$.\\[10pt]
$\mathcal{M}_{6}$ & BMS: DMS with $\alpha=1$, $\delta=1$ and $\beta = 1$.\\[10pt]

$\mathcal{M}_{7}$ & \parbox{14cm}{Kitchen Sink: The constant term, $y_{t-1},\dots,y_{t-4}$ and all exogenous predictors are always included.  We set $\alpha=0.99$, $\beta = 0.96$ and average only over $\delta_{1},\dots,\delta_{d}$.}\\
\hline
\end{tabular}
\caption{Model specifications. The first column is the model index. The second column provides a brief description of each individual model.}
\label{tab:models}
\end{table}

\subsection[Out--of--sample forecasts]{Out--of--sample forecasts}

An important feature of DMA is out--of--sample forecasting,
see \cite{koop_korobilis.2011} and \cite{koop_korobilis.2012}. In this
section, we illustrate how our package can be used to generate forecasts.

In Table~\ref{tab:models}, we provide an overview of several alternative models.
Notice that, all models can be estimate using our package. For instance, the plain AR(4) model, ($\mathcal{M}_0$), can be estimated by setting $\delta=1.0$, $\alpha=1.0$, $\beta = 1.0$, using the code:

\begin{CodeChunk}
\begin{CodeInput}
R> Fit_M0 <- DMA(GDPDEF ~  Lag(GDPDEF, 1) + Lag(GDPDEF, 2) +
                           Lag(GDPDEF, 3) + Lag(GDPDEF, 4),
                           data = USData, vDelta = 1.00,
                           dAlpha = 1.00, vKeep = c(1, 2, 3, 4, 5),
                           dBeta = 1.0)
\end{CodeInput}
\end{CodeChunk}

Where \code{vKeep = c(1, 2, 3, 4, 5)} indicate that all the predictors are included\footnote{This is equivalent to \code{vKeep = "KS"}.}. The same holds for Bayesian Model Averaging (BMA, $\mathcal{M}_5$) and Bayesian Model Selection (BMS, $\mathcal{M}_6$) by setting
$\delta=1.0$, $\alpha=1.0$ and $\beta = 1.0$. Thus, \pkg{eDMA} also relates to the \pkg{BMS} package of \cite{zeugner_feldkircher.2015} and the \pkg{BMA} package of \cite{BMA}.

We use the models to obtain one ($h = 1$) and five ($h = 5$) quarter ahead forecasts through
direct forecasting, see \cite{marcellino_etal.2006}.

Table~\ref{tab:back} reports the mean squared error (MSE) and the log--predictive likelihood difference (PLD) of
$\mathcal{M}_{i}$, $i=1,\dots,7$, over $\mathcal{M}_{0}$ (the benchmark)
at $h=1$ and $h=5$.\footnote{We recall that multi step ahead forecast is performed via direct forecasting as in \cite{koop_korobilis.2012}. For instance, the \code{formula} used for model $\mathcal{M}_0$ when $h=5$ is \code{GDPDEF $\sim$  Lag(GDPDEF, 5) + Lag(GDPDEF, 6) +  Lag(GDPDEF, 7) + Lag(GDPDEF, 8)}}

Compared to the benchmark, $\mathcal{M}_{1}$ provides gains both in terms of MSE and PLD relative to the benchmark, especially at $h=5$. By averaging over $y_{t-1},...y_{t-4}$ and accounting for parameter instability, we obtain even more gains. DMA using lags of inflation as well as $15$
additional predictors is the top performer, regardless of $h$. Similar to \cite{groen_etal.2013} the exogenous predictors contain enough information besides the lags the improve forecast accuracy. Conversely, DMS is outperformed by the benchmark at $h=1$. This
result is understandable as panel (c) in Figure~\ref{fig:combined} demonstrates that
no individual model or group of model combinations perform overwhelmingly better than
the other specifications. By looking more carefully at DMS results, we find that at $h=1$, DMS produces volatile forecasts at the start and towards the end of the sample, which explains why it is outperformed by the benchmark. This is evident from panel (b) of Figure~\ref{fig:combined}, where we observe notable changes in the number of predictors in the optimal model at the start of the sample, towards and during the Great Recession of 2008.

As previously mentioned, DMA (DMS) with $\alpha=\delta=\beta=1$ correspond
to BMA (BMS). At $h=1$, compared to
the benchmark model, BMA provides improvements in density and point forecasts. Similar to DMS, BMS is outperformed by the benchmark at $h=1$. At both horizons, results confirm that accounting for model uncertainty and parameter instability lead to more out--of--sample gains.

\begin{table}[!t]
  \centering
  \setlength\tabcolsep{28pt}
  \begin{tabular}{lcccc}
  \toprule
Model  & \multicolumn{2}{c}{$h = 1$} & \multicolumn{2}{c}{$h = 5$}\\
  \cmidrule(lr){1-1}\cmidrule(lr){2-3}\cmidrule(lr){4-5}
 & MSE & PLD & MSE & PLD\\
\cmidrule(lr){1-1}\cmidrule(lr){2-3}\cmidrule(lr){4-5}
$\mathcal{M}_{1}$ & 0.998 & 12.444 & 0.815 & 48.445 \\
$\mathcal{M}_{2}$ & 0.964 & 14.416 & 0.728 & 64.008 \\
$\mathcal{M}_{3}$ & 0.938 & 20.561 & 0.704 & 94.399 \\
$\mathcal{M}_{4}$ & 1.155 & -2.701 & 0.844 & 62.227 \\
$\mathcal{M}_{5}$ & 0.985 & 7.234 & 1.138 & 19.368 \\
$\mathcal{M}_{6}$ & 1.096 & -7.543 & 1.308 & -25.294 \\
$\mathcal{M}_{7}$ & 1.839 & -9.899 & 0.965 & 47.832 \\
\bottomrule
  \end{tabular}
    \caption{Mean squared error (MSE) and log--predictive likelihood difference (PLD) of $\mathcal{M}_{i},\quad i=1,\dots,7$ compared to $\mathcal{M}_{0}$ for $h = 1$ and $h = 5$ quarters ahead out--of--sample forecasts.}
        \label{tab:back}
\end{table}

Finally, as an alternative to these models, we can consider the Kitchen
Sink model (the model with all predictors, $\mathcal{M}_7$) where we only average over $\delta$. Compared to $\mathcal{M}_{0}$, the kitchen
sink model does not provide any improvements at $h=1$. At $h=5$,
we observe improvements in density forecasts compared to $\mathcal{M}_0$. However, the kitchen
sink model is always outperformed by DMA.

\subsection[Why does DMA perform well ?]{Why does DMA perform well ?}

To investigate how quickly our techniques adapt to changes in data,
we report the accumulated log--PLD for several models over the benchmark
in panels (a)--(d) of Figure~\ref{fig:pld_size}. These can be obtained using the \code{pred.like()} method available for \code{DMA} objects. For instance, we create the two vectors \code{vPL_M0} and \code{vPL_M3} containing the log--predictive likelihood of $\mathcal{M}_0$ and $\mathcal{M}_3$ using:

\begin{figure}[!t]
\centering
\includegraphics[width=1\textwidth]{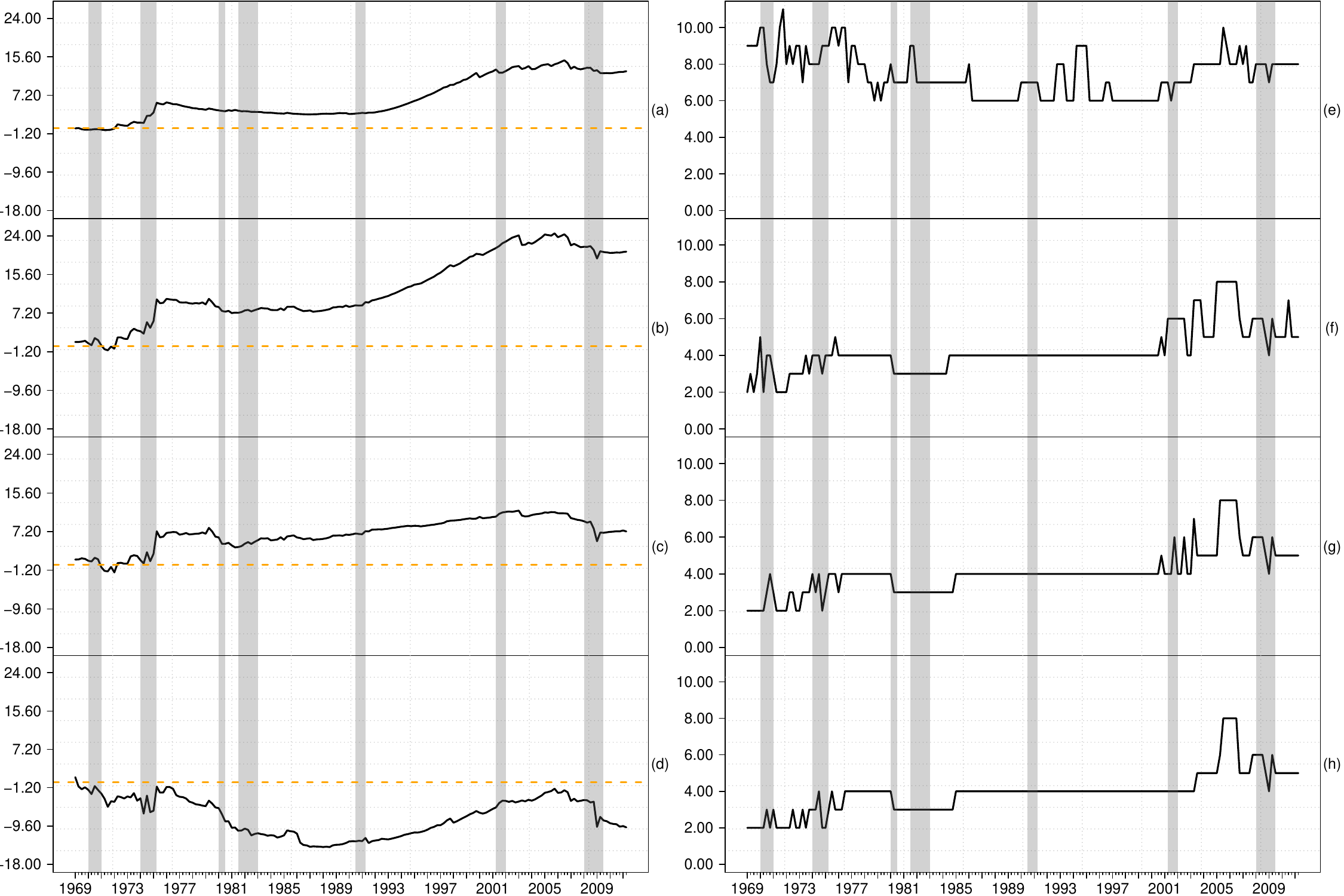}
\caption{Accumulated PDL and the optimal number of predictors for prior sensitivity analysis. Panel (a): $\mathcal{M}_1$ over $\mathcal{M}_0$, Panel (b): $\mathcal{M}_3$ over $\mathcal{M}_0$, Panel (c): $\mathcal{M}_5$ over $\mathcal{M}_0$, Panel (d): $\mathcal{M}_7$ over $\mathcal{M}_0$. Panels (e)--(h): Number of predictors for the model with the highest posterior probability using $g = 0.1, 20, T/2, T$. The gray vertical bars indicate business cycle peaks, \emph{i.e.}, the point at which an economic expansion transitions to a recession, based on National Bureau of Economic Research (NBER) business cycle dating.}
\label{fig:pld_size}
\end{figure}

\begin{CodeChunk}
\begin{CodeInput}
R> vPL_M0 <- pred.like(Fit_M0, iBurnPeriod = 32)
R> vPL_M3 <- pred.like(Fit, iBurnPeriod = 32)
\end{CodeInput}
\end{CodeChunk}

and compute the accumulated log-PLD of $\mathcal{M}_3$ over $\mathcal{M}_0$ as:
\begin{CodeChunk}
\begin{CodeInput}
R> vPLD_M3.M0  <- cumsum(vPL_M3 - vPL_M0)
\end{CodeInput}
\end{CodeChunk}

which is reported in panel (b) of Figure~\ref{fig:pld_size}.

In panels (a), (b), (c) and (d) of Figure~\ref{fig:pld_size} a value of zero corresponds to equal support of both
models, positive values are in support of the model of choice over
$\mathcal{M}_{0}$ and negative values show support of $\mathcal{M}_{0}$
over the model of choice at time $t$. In these panels, we decompose the effects
of (i): Allowing for time--variation in the regression coefficients,
(ii): Allowing for model uncertainty but no time--variation in the
regression coefficients and (iii): Allowing for time--variation in
the regression coefficients and model uncertainty.

In panel (a), we see that the time--varying AR(4) model outperforms the benchmark throughout the out--of--sample period. Compared to the plain AR(4) model, it takes about twenty observations
to provide compelling evidence in favor of DMA. Furthermore, we also observe that DMA performs well in recession as well as expansion periods. Compared
to BMA, the improvements of DMA are mostly concentrated on the onset of recessions. However, DMA also outperforms BMA during expansion periods. Conversely the kitchen sink model is generally outperformed by the benchmark throughout the out--of--sample, see panel (d) of Figure~\ref{fig:pld_size}.

\subsection[The choice of g]{The choice of $g$}
\label{sec:sensitivity}

In the context of DMA, the prior hyperparameter value, $g$, must
be specified by the practitioner. Intuitively, a smaller value of
$g$ means more shrinkage around the prior mean of $\btheta^{(i)}_{0}$,
\emph{i.e.}, $\boldsymbol{0}$. The larger is $g$, the more we are willing to move away
from the model priors in response to what we observe in the data.
In other words, the larger the $g$, the more we allow data to speak
freely. This way, we ensure that the estimation
procedure quickly adapts to data, even at quarterly frequency, which
typically consist of around $300$ observations. On the other hand,
for some data--sets, it can take the estimation procedure longer time to adapt if
we set $g$ to relatively lower values. Thus, in such cases, DMA can initially overfit as the average model size becomes larger than
it ought to be. This effect becomes evident by examining the
average number of predictors in DMA and in most cases is also heavily
reflected in the generated forecasts, where DMA is outperformed
by the benchmark.

We re--estimate DMA with $g$ equal to $0.1$,
$20$, $T/2$ and $T$ (using \code{bZellnerPrior = FALSE}) and observe to which extent different values of $g$ influences out--of--sample
results. Results are reported in Table~\ref{tab:priorsensitivity} and panels (e)--(h) of Figure~\ref{fig:pld_size}.
Overall, we find that results are robust to different values of $g$. All $g$ values lead to similar MSE and PLD estimates and the number of predictors in the model with the highest posterior probabilities are also similar, see panels (e)--(h) of Figure~\ref{fig:pld_size}. However, we must mention that this is mainly due to the properties of our data and the fact that \code{bZellnerPrior = FALSE} such that, contrary to \code{bZellnerPrior = TRUE}, the observations do not affect the prior covariance matrix of $\btheta^{(i)}_t$, see Equation \ref{eq:prior}. In fact, when we repeat the analysis with \code{bZellnerPrior = TRUE}, we find that DMA using $g=0.1$ and $g=20$ perform much worse and are outperformed by the benchmark model. On the other hand, as we increase $g$ to $T/2$ and $T$, we obtain similar results to those reported in Table~\ref{tab:priorsensitivity}. This result is understandable as given the scale of the prior covariance matrix under \code{bZellnerPrior = TRUE}, prior shrinkage is much greater under $g=0.1$ and $g=20$.

Ultimately, it is up to the practitioner to choose $g$. However, our general recommendation is to fix $g=T$ regardless of \code{bZellnerPrior = TRUE} or \code{FALSE} and the number of observations as it allows the data to speak freely about the underlying relations between the regressors and the dependent variable. However, as previously mentioned, we recommend \code{bZellnerPrior = FALSE}, for small data-sets.\footnote{An anonymous referee also makes a very good point regarding choosing $g$, which can be summarized as follows:

\begin{itemize}
\item[(i):] Choose $b$ values of $g$, say $g=\left\{0.1,20,T/2,T\right\} $.
Then run DMA for each of these values and save the predictive likelihoods
$p\left(y_{t}^{\left(i\right)}\mid\mathcal{F}_{t-1}\right)$, $t=1,...,T$
for $i=1,\dots,b$

\item[(ii):] Compute $p\left(y_{t}^{\left(i\right)}\vert\mathcal{F}_{t-1}\right)/\Sigma_{i=1}^{b}p\left(y_{t}^{\left(i\right)}\vert\mathcal{F}_{t-1}\right)$,
$t=1,\dots,T$ for $i=1,\dots,b$.

Thus, we can observe which value of $g$ obtains high posterior
probabilities, especially at the start of the sample. We can then
use the associated $g$ value in the estimation procedure.
\end{itemize}}

\begin{table}[!t]
  \centering
  \setlength\tabcolsep{55pt}
  \begin{tabular}{lcc}
  \toprule
Prior & MSE & PLD \\
\cmidrule(lr){1-1}\cmidrule(lr){2-3}
$g = 0.1$ & 0.967 & 20.186 \\
$g = 20$ & 0.937 & 20.664 \\
$g = T/2$ & 0.938 & 20.556 \\
$g = T$ & 0.941 & 20.431 \\
\bottomrule
  \end{tabular}
    \caption{Mean squared error (MSE) and log--predictive likelihood difference (PLD) of DMA using the following values of $g$: $0.1,20,T/2,T$ and $\mathcal{M}_{0}$ for $h = 1$.}
     \label{tab:priorsensitivity}
\end{table}

\section[Conclusion]{Conclusion}
\label{sec:conclusion}
In this paper, we present the \pkg{eDMA} package for \proglang{R}. The purpose of \pkg{eDMA} is to offer an integrated environment to easily perform DMA using the available \code{DMA()} function, which enables practitioners to perform DMA exploiting multiple processors. Furthermore, \proglang{R} users will find common methods to represent and extract estimated quantities such as \code{plot()}, \code{as.data.frame()}, \code{coef()} and \code{residuals()}.

Overall, \pkg{eDMA} is able to: (i): Incorporate the extensions introduced in \cite{prado_west.2010} and \cite{dangl_halling.2012}, which are relevant for economic and financial applications, (ii): Compared to other approaches, our package is much faster, (iii): It requires a smaller amount of RAM even in cases of moderately large applications, and (iv): It allows for parallel computing.

In Section \ref{sec:computational}, we also detail the expected time the program takes to perform DMA under different sample sizes, number of predictors and number of grid points. For typical economic applications, estimation time is around 30 minutes using a commercial laptop. Large applications can still benefit from the use of \pkg{eDMA} even when performed on desktop or clusters, without additional effort from the user. 

\section*{Computational details}

The results in this paper are obtained using \proglang{R} 3.2.3 \citep{R.2015} with the
packages: \pkg{eDMA} version 1.4-0 \citep{eDMA},
\pkg{Rcpp} version 0.12.5 \citep{Rcpp.2011,Rcpp}, \pkg{RcppArmadillo} version 0.7.100.3.1 \citep{RcppArmadillo.2014,RcppArmadillo}, \pkg{xts} version 0.9-7 \citep{xts} and \pkg{devtools} version 1.1.1 \citep{devtools}.
\proglang{R} itself and all packages used are available
from \proglang{CRAN} at \url{http://CRAN.R-project.org/}. The package \pkg{eDMA} is available from CRAN at \url{https://cran.r-project.org/web/packages/eDMA/index.html}.
Computations were performed on
a Genuine Intel\textregistered{} quad core CPU i7--3630QM 2.40Ghz processor.

\clearpage


\clearpage

\appendix

\section[The mathematics of dynamic linear models]{The mathematics of dynamic linear models}
\label{sec:app1}

Below, we briefly outline the Kalman recursions for the $i$-th DLM model in the model average. We can refer the reader to \cite{prado_west.2010} for more details on the Kalman recursions. Based on the information up to time $t-1$, the prior if the state vector, $\btheta_{t}^{\left(i\right)}$, at time $t$ follows $\mathcal{N}\left(\ba_{t}^{\left(i\right)},\bR_{t}^{\left(i\right)}\right)$, where:

\begin{align}
\ba_{t}^{\left(i\right)} &= \bm_{t-1}^{\left(i\right)},\nonumber\\
\bR_{t}^{\left(i\right)} &= \bC_{t-1}^{\left(i\right)} + \bW_{t}^{\left(i\right)}.\label{eq:A.1}
\end{align}

Conditional on $V_{t}^{(i)}$, the one--step--ahead predictive mean and variance
of $y_{t}^{\left(i\right)}$ follows a Normal distribution
with mean, $f_{t}^{\left(i\right)}$, and variance, $Q_{t}^{\left(i\right)}$,
where:

\begin{align}
f_{t}^{\left(i\right)} &= \bF_{t}^{\left(i\right)^\prime}\ba_{t}^{\left(i\right)}\nonumber,\\ Q_{t}^{\left(i\right)} &=  \bF_{t}^{\left(i\right)^\prime}\bR_{t}^{\left(i\right)}\bF_{t}^{\left(i\right)} + V_{t}^{\left(i\right)}.\label{eq:A.2}
\end{align}

Once we observe $y_{t}$, we can compute the forecast error as $e_{t}^{\left(i\right)}=y_{t}-f_{t}^{\left(i\right)}$.
The posterior distribution for $\btheta_{t}^{\left(i\right)}$
given the current information set, $\mathcal{F}_{t}$, is then updated
as:

\begin{align}
\bm_{t}^{\left(i\right)} &= \ba_{t}^{\left(i\right)} + \bA_{t}^{\left(i\right)}e_{t}^{\left(i\right)}\nonumber,\\
\mathbf{C}_{t}^{\left(i\right)} &=  \bR_{t}^{\left(i\right)}-\bA_{t}^{\left(i\right)}\bA_{t}^{\left(i\right)^\prime}Q_{t}^{\left(i\right)},\label{eq:A.3}
\end{align}

where $\bA^{(i)}_t$ is the adaptive coefficient vector $\bA_{t}^{\left(i\right)} = \bR_{t}^{\left(i\right)}\bF_{t}^{\left(i\right)}/Q_{t}^{\left(i\right)}$.

\section[True out--of--sample forecast]{True out--of--sample forecast}
\label{sec:app2}

There might be cases where the practitioner desires to predict $T+1$ conditional on observation till time $T$ in a true out--of--sample fashion (\emph{i.e.}, without having the possibility of backtesting the forecast since $y_{T+1}$ cannot be observed). In such circumstances, the user can substitute the future value of the dependent variable with an \code{NA}. This way, the code treats the last observation as missing and does not perform backtesting or updating of the coefficients. However, the estimation procedure provides us with the necessary quantities to perform prediction. The predicted value $\widehat{y}_{T+1} = \E[y_{T+1}\vert\mathcal{F}_T]$ as well as the predicted variance decomposition defined in Equation \ref{eq:varDec} can then be extracted using the \code{getLastForecast} method available in the \pkg{eDMA} package. The other quantities that can be extracted, for example via the \code{as.data.frame} method, will ignore the presence of the last \code{NA} and report results only for the firs+t $T$ observations.

For example, consider the simulated data, \code{SimData}, detailed in Section \ref{sec:using}

\begin{CodeChunk}
\begin{CodeInput}
R> data("SimData", package = "eDMA")
\end{CodeInput}
\end{CodeChunk}

Recall that this is a $500\times 6$ \code{dataframe} simulated from the model defined in Equations \ref{eq:dgp} - \ref{eq:theta_rw}. The first column represents the dependent variable, $y_t$, while the last five columns the predictors $x_{i,t}$ for $i = 2,\dots,6$ and $t = 1,\dots,T=500$. Assume that we observe (or that we have previously forecasted) the values for the predictors at time $T+1$, \emph{i.e.}, $x_{i, T+1}\in\mathcal{F}_T$ for $i = 2,\dots,6$, and these are $x_{2,T+1} = -0.07$, $x_{3,T+1} = 1.61$, $x_{4,T+1} = -2.07$, $x_{5,T+1} = 0.17$, $x_{6,T+1} = -0.80$. What we need to do it is simply bind a new row at the \code{SimData} \code{dataframe}

\begin{CodeChunk}
\begin{CodeInput}
R> newData <- c(NA, -0.07, 1.61, -2.07, 0.17, -0.80)
R> SimData <- rbind(SimData, newData)
\end{CodeInput}
\end{CodeChunk}

and run DMA

\begin{CodeChunk}
\begin{CodeInput}
R> Fit <- DMA(y ~ x2 + x3 + x4 + x5 + x6 , data = SimData,
          vDelta = seq(0.9, 1.0, 0.01))
\end{CodeInput}
\end{CodeChunk}

In order to extract the predicted value $\widehat{y}_{T+1} = \E[y_{T+1}\vert\mathcal{F}_T]$ and the predicted variance decomposition, we simply run

\begin{CodeChunk}
\begin{CodeInput}
R> getLastForecast(Fit)
\end{CodeInput}
\begin{CodeOutput}
$PointForecast
[1] 11.5293

$VarianceDecomposition
      vtotal         vobs       vcoeff         vmod         vtvp
4.290887e-01 2.805227e-01 1.478767e-01 6.682507e-04 2.108273e-05
\end{CodeOutput}
\end{CodeChunk}

\clearpage

\end{document}